\documentclass[11pt,a4paper]{article}
\usepackage[margin=25mm]{geometry}
\usepackage{setspace}
\usepackage[T1]{fontenc}
\usepackage[utf8]{inputenc}
\usepackage{amsmath,amssymb,amsfonts,bm}
\usepackage{graphicx}
\usepackage{braket}
\usepackage{placeins}
\usepackage{caption}
\usepackage[super,sort&compress,numbers]{natbib}

\usepackage{xurl}
\usepackage{hyperref}
\hypersetup{colorlinks=true,linkcolor=black,citecolor=black,urlcolor=black}
\begin{document}
\begin{center}
{\LARGE\bfseries{The noise cost of reservoir locality in broadband quantum amplifier chains}\par}
\vspace{0.7em}
Chengxi Li$^{1,2}$, Hao Zhu$^{1,2}$, Tie-Fu Zhang$^{2,3}$, Lin Zhuang$^4$,
Wanzi Sun$^{2,5,*}$ and Wuming Liu$^{2,*}$\par
\vspace{0.7em}
\end{center}
{\small
$^1${Beijing Institute of Petrochemical Technology, Beijing 102617, China}\par
$^2${Beijing National Laboratory for Condensed Matter Physics, Institute of Physics, Chinese Academy of Sciences, Beijing 100190, China}\par
$^3${Faculty of Science, Kunming University of Science and Technology, Kunming 650500, China}\par
$^4${State Key Laboratory of Optoelectronic Materials and Technologies, School of Physics, Sun Yat-Sen University, Guangzhou, 510275, China.}\par
$^5${Jining University, Qufu, Shandong 273155, China}\par
\medskip
$^*$Correspondence: Wanzi Sun
(\href{mailto:wzsun@jnxy.edu.cn}{wzsun@jnxy.edu.cn}) and Wuming Liu
(\href{mailto:wmliu@iphy.ac.cn}{wmliu@iphy.ac.cn}).\par}

\section*{Abstract}
Directional quantum amplifiers enable weak-signal readout while
protecting the source, but reservoir noise limits sensitivity.
Finite-band suppression must reconcile frequency-dependent responses
with fixed local couplings and finite rates.  We construct adjacent-site
loss channels attaining the known single-frequency minimum under local
compatibility conditions, as an uncapped benchmark.
For specified uniform and dimerized chains, we compare fixed reservoirs
at matched drift, ports, high-gain band and complete gain/loss rate caps.
At the representative cap $R=8$ and halfwidth $\Omega=0.05$, nearest-neighbor arithmetic
band-average noise exceeds the unrestricted optimum by more than
$1.01265$ quanta in the
uniform chain for every finite $N\ge64$.  Boundary-response bounds give
necessary rates and targets requiring range two.  At $N=64$,
finite-auxiliary calculations retain a noise advantage under matched
total coupling budgets; response bounds quantify noise-resolution
tolerances.  These theoretical results guide reservoir design and noise calibration
for a proposed cryogenic microwave array with a $100$ kHz design band.

\section{Introduction}

Quantum amplifiers convert weak signals into measurable outputs and
are central to cryogenic microwave readout \cite{Clerk2010}.
Directional amplification combines forward gain with suppressed reverse
transmission, protecting the source from the readout circuit
\cite{MetelmannClerk2015,MetelmannClerk2017,PorrasFernandez2019,Wanjura2020,RamosDirectional2021,LiuStable2023}.
In driven bosonic arrays, interference between coherent hopping,
parametric driving and engineered dissipation produces this response
\cite{PeanoEtAl2016,WangEtAl2022NHSEAmplification,WanjuraDisorder2021,GomezLeonEtAl2023}.
These arrays connect amplifier design to non-Hermitian lattice physics
\cite{AshidaGongUeda2020,BergholtzBudichKunst2021}, where boundary
localization \cite{HatanoNelson1996,YaoWang2018,Okuma2020,GuoAnomalousSkin2023}
and exceptional points \cite{ZhangLieb2023} alter finite-device response.
Electrical and photonic lattices provide experimental realizations of
non-Hermitian propagation \cite{Helbig2020,Weidemann2020}.

Gain alone does not determine sensitivity: reservoir fluctuations can
mask a weak input signal \cite{HausMullen1962,Caves1982,CavesCombes2012}.
Reservoir engineering enables quantum-state preparation \cite{Diehl2008}
and quantum-limited amplification \cite{MetelmannClerk2014}, but identical
linear signal dynamics can
arise from reservoirs with different output noise
\cite{LauClerk2018,HernandezLiberal2024}.
Lau and Clerk obtained the fixed-drift single-frequency minimum and
noted that one bath need not optimize multiple detunings
\cite{LauClerk2018}.  A physical amplifier must instead use one set of
couplings across the band, with finite range and finite rates.
Long-range dissipation can produce multimode topological amplification
by reshaping the drift \cite{VegaEtAl2025}.
At a prescribed signal response, it remains unresolved whether longer
chains can compensate for local reservoir couplings, or whether a
broadband noise cost persists when the detection ports and complete
gain/loss budgets are also held fixed.
Input-stage noise dominance in independent amplifier cascades
\cite{Friis1944} does not resolve this question: shared reservoirs can
correlate across that region while gain and loss remain tied by the
fixed drift.

Here, adjacent-site loss couplings orthogonal to the output response
attain the known pointwise minimum under local compatibility conditions,
without a prescribed rate cap.
For fixed broadband reservoirs in the specified uniform and dimerized
chains, an input-boundary region retains the same input-normalized
noise weights as the chain grows.  This gives a positive nearest-neighbor
penalty exceeding $1.01265$ input-referred quanta in the uniform chain
for every finite $N\ge64$ at rate cap $R=8$ and halfwidth $\Omega=0.05$.
The bounds give necessary rates and noise targets requiring range two,
which recovers $91.6\%$ of the nearest-neighbor versus unrestricted gap
at $N=64$.
Finite-auxiliary calculations at $N=64$ test the advantage under matched
total coupling budgets; response bounds determine the uncertainty
allowed in resolving the noise difference.

\begin{figure}[htbp]
\centering
\includegraphics[width=\textwidth]{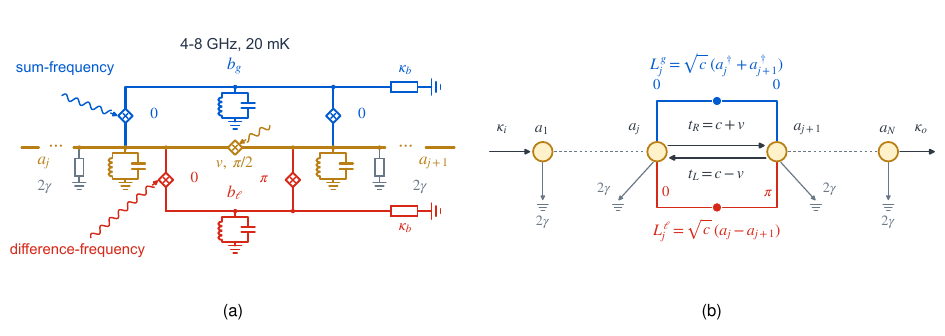}
\caption{Baseline reservoir circuit and effective directional chain.
(a) Proposed implementation of one bond: gold signal LC modes couple
through Josephson mixers (diamonds)
to blue/red gain/loss auxiliaries $b_g,b_\ell$, independently terminated
at energy linewidth $\kappa_b$. The gold link provides coherent conversion;
phase labels refer to calibrated couplings. LC grounds are electrical references.
Representative pumps are shown; repeated units and readout are omitted.
(b) Drift arrows $t_R=c+v$, $t_L=c-v$ combine coherent hopping and
the displayed jumps, with gain phases $0,0$ and loss phases $0,\pi$.
Gray loads and sinks denote on-site energy loss $2\gamma$; rates are
nonnegative. Optimization reprograms couplings at fixed drift and ports;
range-two reservoirs address up to three consecutive signal sites.}
\label{fig:device}
\end{figure}

Our experimental setup would use a superconducting microwave array
of $4$--$8$ GHz resonators at $20$ mK, coupled by parametrically driven
Josephson mixers [Fig.~\ref{fig:device}(a)] \cite{Sliwa2015,Lecocq2017}.
Below, rates and detunings are dimensionless in units of a reference
angular frequency $J_0$, with time in $J_0^{-1}$.
At $J_0/(2\pi)=1$ MHz, the proposed $N=64$ design uses
coherent hopping $J_0v/(2\pi)=0.375$ MHz, bond-reservoir rate
$J_0c/(2\pi)=0.625$ MHz, and input and output port energy linewidths
$J_0\kappa_i/(2\pi)=J_0\kappa_o/(2\pi)=0.2$ MHz.
The detection band is $\pm50$ kHz, with full width $100$ kHz.
Auxiliaries with independent cold baths and matched energy linewidth
$J_0\kappa_b/(2\pi)=80$--$100$ MHz mediate correlated gain and loss, with baseline
coupling $J_0|g|/(2\pi)=3.54$--$3.95$ MHz per signal mode.
Phase-locked pumps \cite{RanzaniAumentado2015} remain fixed over the band.
Passive end waveguides provide coherent probes and quadrature readout.
Eliminating the rapidly damped auxiliaries leaves the effective signal
model below.

\FloatBarrier

\section{Results}

\subsection{Physical model and measured noise}
The resulting open-chain model has $N$ signal resonators, with
$a_j$ ($a_j^\dagger$) annihilating (creating) a photon at site $j$.
The passive input port couples to the first site and the output port
to the last; their site indices are denoted by $i=1$ and $o=N$,
respectively [Fig.~\ref{fig:device}(b)].
In frames rotating at the resonator frequencies, the resonant conversion
links generate number-conserving hopping in the rotating-wave approximation.
For a conversion phase $\pi/2$, the Hermitian signal Hamiltonian is
\begin{equation}
\widehat H={\mathrm{i}} v\sum_{j=1}^{N-1}(a_{j+1}^\dagger a_j-a_j^\dagger a_{j+1}),
\label{eq:signal_hamiltonian_nc}
\end{equation}
where $v=(t_R-t_L)/2$ is the coherent hopping strength, while $t_R$ and $t_L$
are the rightward and leftward couplings in the amplitude dynamics
derived below.  We set $\hbar=1$.

The difference-frequency pump couples the loss auxiliary to
$a_j-a_{j+1}$, while the sum-frequency pump couples the gain auxiliary
to $a_j^\dagger+a_{j+1}^\dagger$, with complex coupling amplitude $g$
per signal mode.  Cold terminations give each auxiliary an energy decay
rate $\kappa_b$ and hence amplitude decay rate $\kappa_b/2$.
The probe detuning $\omega$ is measured from the signal carrier.
For resonant pumps and $|g|,|\omega|\ll\kappa_b$, the auxiliaries relax
rapidly and can be eliminated, giving the bond-reservoir rate
$c=4|g|^2/\kappa_b$ to leading Markov order
\cite{GardinerZoller2004,MetelmannClerk2015}.
The bond jumps are
\begin{equation}
\begin{aligned}
L_j^g&=\sqrt c(a_j^\dagger+a_{j+1}^\dagger),\\
L_j^\ell&=\sqrt c(a_j-a_{j+1}),
\end{aligned}
\label{eq:bond_jumps_nc}
\end{equation}
with $c=(t_R+t_L)/2$.
On-site energy loss $2\gamma$, with $\gamma$ the amplitude decay rate,
is described by $L_j^0=\sqrt{2\gamma}\,a_j$.
The superscripts $g,\ell,0$ label correlated gain, correlated loss, and
on-site loss, respectively; $c$ is the bond-reservoir rate.
\par
These dissipative couplings interfere with coherent hopping to produce
nonreciprocal signal dynamics.  The signal density operator $\rho$ obeys
the Lindblad master equation
\cite{Lindblad1976,GKS1976}
\begin{equation}
\dot\rho=-{\mathrm{i}}[\widehat H,\rho]+\sum_\mu\mathcal D[L_\mu]\rho,
\label{eq:lindblad_master_prl}
\end{equation}
where $\mathcal D[L]\rho=L\rho L^\dagger-\{L^\dagger L,\rho\}/2$ and
$\mu$ labels reservoir channels.
For the mode vector $\bm a=(a_1,\ldots,a_N)^{\mathsf{T}}$, this gives
${\mathrm{d}}\langle\bm a\rangle/{\mathrm{d}} t=X\langle\bm a\rangle$, where $X$ is the
drift matrix governing the mean signal amplitudes, with entries
$X_{jj}=-\gamma$, $X_{j+1,j}=t_R$, and $X_{j,j+1}=t_L$.
Here $t_R=c+v$ and $t_L=c-v$ combine dissipative and coherent couplings.
We use the dimensionless parameters $t_R=1$, $t_L=0.25$, and $\gamma=1.2$, within the stable
directional-amplification window
$2\sqrt{t_Rt_L}<\gamma<t_R+t_L$.
This response window follows from the endpoint Green function
\cite{YokomizoMurakami2019,XueDirectional2020,LiWan2021}.
Damping gaps
\cite{ProsenSeligman2010,BarthelZhang2022}, finite-chain stability
\cite{UghrelidzeEtAl2024}, and non-normal relaxation
\cite{TrefethenEmbree2005,Haga2021,MoriShirai2020,MoriShirai2023}
are distinct dynamical properties.
The bounds below apply to stable Markov dynamics; finite-linewidth
corrections are derived in Sec.~\mbox{S-5\,A} of the Supplementary Information~\cite{SupplementalMaterial}.

For noise optimization, keep this signal drift fixed and allow general linear channels
$L_\mu^g=\bm a^\dagger g_\mu$ and $L_\mu^\ell=\ell_\mu^\dagger\bm a$.
Here $g_\mu$ and $\ell_\mu$ collect the complex jump amplitudes over the sites for gain
and loss channel $\mu$, respectively.  Their positive semidefinite rate matrices are
$\Gamma_g=\sum_\mu g_\mu g_\mu^\dagger$ and
$\Gamma_\ell=\sum_\mu\ell_\mu\ell_\mu^\dagger$.
The drift relation is
\begin{equation}
X=-{\mathrm{i}} H+(\Gamma_g-\Gamma_\ell)/2,
\label{eq:drift_relation_prl}
\end{equation}
where $H$ is the single-particle hopping matrix, $\widehat H=\bm a^\dagger H\bm a$.
Fixed drift therefore requires $\Gamma_g-\Gamma_\ell=M\equiv X+X^\dagger$,
with $M$ the fixed net gain-minus-loss matrix.
Different positive semidefinite rate matrices with the same difference
preserve coherent transmission while changing the reservoir noise.

To determine the measured output, include the passive ports.
Let $e_j$ denote the unit vector selecting site $j$.
The port energy linewidths $\kappa_i=\kappa_o=0.2$ add boundary damping, giving
$X_\kappa=X-(\kappa_i e_i e_i^\dagger+\kappa_o e_o e_o^\dagger)/2$.
The transmission response to a coherent probe follows input-output theory
\cite{GardinerCollett1985,GardinerZoller2004} and is described by the susceptibility matrix
$\chi_\kappa(\omega)=({\mathrm{i}}\omega I-X_\kappa)^{-1}$, with $I$ the identity matrix.
Define the output-response vector $y=\chi_\kappa^\dagger e_o$:
$y_j^*$ is the propagation amplitude from a fluctuation at site $j$ to
the output.  The power gain is $G(\omega)=\kappa_i\kappa_o|y_i|^2$.
Dividing the normally ordered output photon noise above vacuum by gain
gives the input-referred noise
\begin{equation}
\mathcal N(\omega)=\frac{y^\dagger\Gamma_g y}{\kappa_i|y_i|^2}.
\label{eq:input_referred_noise_nc}
\end{equation}
Its numerator projects gain-reservoir noise onto
the measured output response.  All comparisons include the exact port damping and
assume stable $X_\kappa$ and $y_i\ne0$.
For symmetric spectra, the conventional symmetrized added noise is
$n_{\rm add}=\mathcal N-(1-G^{-1})/2$; its fixed-drift differences coincide
with ours, as derived in Sec.~\mbox{S-1\,B} of the Supplementary
Information~\cite{SupplementalMaterial}.

\FloatBarrier

\subsection{Pointwise local optimality}
The fixed-drift constraint and positivity of both rate matrices give
$\mathcal N(\omega)\ge f(\omega)$, with the known lower bound
\cite{LauClerk2018}
\begin{equation}
f(\omega)=\frac{\max\{y^\dagger My,0\}}{\kappa_i|y_i|^2}.
\label{eq:pointwise_floor_nc}
\end{equation}
For $y^\dagger My>0$, the excess is
$\mathcal N-f=y^\dagger\Gamma_\ell y/(\kappa_i|y_i|^2)$.
Vacuum loss baths add no normally ordered photons; this excess is the
additional gain noise required to compensate loss at fixed drift.
For a two-site loss coupling, the two response amplitudes cancel when
its coupling vector is orthogonal to $y$.  The excess is a sum of
nonnegative squared overlaps, so each channel must satisfy this condition.
The balanced uniform-chain construction approaches a complete rate
of $139.5J_0$, exceeding $R=8$ and the proposed auxiliary linewidth;
pointwise attainment is an uncapped benchmark, not a setting of the proposed device.

For one signal mode per site with every $y_j\ne0$, define the adjacent-site loss-channel vectors
\begin{equation}
v_j=\frac{y_{j+1}^*e_j-y_j^*e_{j+1}}{\sqrt{|y_jy_{j+1}|}}.
\label{eq:orthogonal_loss_nc}
\end{equation}
These build a positive semidefinite nearest-neighbor loss matrix
$K=\sum_jv_jv_j^\dagger$ with $Ky=0$ and
$\ker K=\operatorname{span}\{y\}$.
If $y^\dagger My>0$, choose
\begin{equation}
\Gamma_\ell=\alpha K,\qquad \Gamma_g=M+\alpha K.
\label{eq:local_construction_prl}
\end{equation}
Here $\alpha>0$ sets the loss strength.  A finite, sufficiently large $\alpha$ makes the gain matrix positive semidefinite,
while $\Gamma_\ell y=0$ gives $\mathcal N=f$.
The positive semidefinite tridiagonal rate matrices in
Eq.~\eqref{eq:local_construction_prl} admit jump factorizations supported
on one or two adjacent sites, giving a local physical realization.
For block-tridiagonal drifts, the construction extends to nonzero
cell-response blocks; exchanging gain and loss treats $y^\dagger My<0$,
as proved in Secs.~\mbox{S-2\,A,C} of the Supplementary Information~\cite{SupplementalMaterial}.
At $y^\dagger My=0$, finite-rate attainment requires $My=0$;
otherwise zero noise is only an infimum.

\begin{figure}[htbp]
\centering
\includegraphics[width=\textwidth]{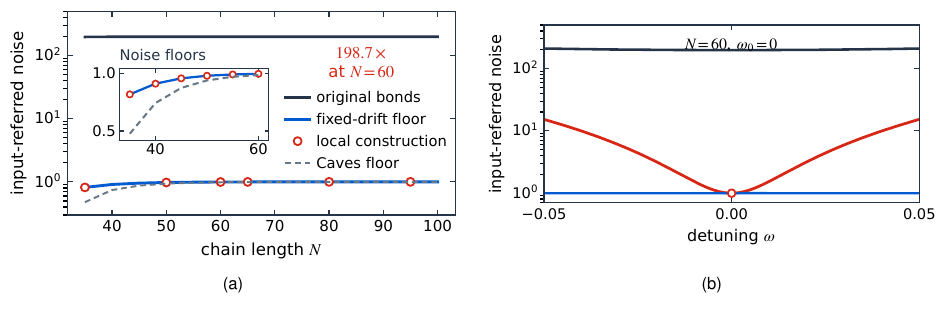}
\caption{Pointwise suppression and its frequency dependence.
(a) Input-referred noise $\mathcal N$, the normally ordered excess output photon noise
divided by power gain $G$, versus chain length $N$ at $\omega=0$:
original bonds (black), local construction (red circles), fixed-drift
minimum $f$ (blue), and Caves floor $1-G^{-1}$ (dashed).
The inset compares the fixed-drift and Caves floors, whose difference
is the reflection penalty $|S_{oo}|^2/G$;
at $N=60$, the original-to-local noise ratio is $198.7$.
(b) Frequency scan of that same $N=60$ reservoir, tuned at zero and then
held fixed (red); black and blue show the original reservoir and
pointwise floor.  Both panels use $t_R=1$, $t_L=0.25$, $\gamma=1.2$,
and $\kappa_i=\kappa_o=0.2$, with logarithmic ordinates.
Both panels use the balanced kernel in Eq.~\eqref{eq:local_construction_prl},
tuned at $\omega_0=0$.  At $N=60$, the complete gain/loss rates are
$134.6J_0$/$139.5J_0$, without a shared cap.  Drift, ports and gain match.}
\label{fig:pointwise_noise}
\end{figure}

Figure~\ref{fig:pointwise_noise}(a) shows noise reduction from $197.658$
to $0.994686$ at $N=60$, $\omega=0$, and unchanged power gain $G=65.894$,
giving $\mathcal N_{\rm original}/\mathcal N_{\rm local}\simeq198.7$.
The residual $0.009862$ above the Caves floor $0.984824$ is the reflection
penalty $|S_{oo}|^2/G$, where $S_{oo}$ is the output-port reflection amplitude.
Figure~\ref{fig:pointwise_noise}(b) holds that particular reservoir fixed:
detuning restores excess gain noise as the response rotates.
This comparison matches drift and ports, without a common rate cap;
it illustrates why pointwise optimality must be followed by a shared-band test.

\FloatBarrier

\subsection{Broadband optimization of shared reservoirs}
The cancellation condition depends on $y(\omega)$.
Across a band, this response direction changes, whereas a fixed device
uses one set of gain/loss couplings.  Positivity then limits the freedom
to suppress all contributing directions.  To compare the resulting
devices fairly, fix the detection band $\mathcal B=[-\Omega,\Omega]$ of halfwidth $\Omega$.
For one reservoir pair held fixed across the band, impose
\begin{equation}
\begin{aligned}
\Gamma_g-\Gamma_\ell&=M,\\
0\preceq\Gamma_g,\Gamma_\ell&\preceq RI,\\
(\Gamma_g)_{jk}=(\Gamma_\ell)_{jk}&=0\quad (|j-k|>r).
\end{aligned}
\label{eq:reservoir_constraints_nc}
\end{equation}
The integer $r$ measures distance in physical sites; $r=1$ is nearest
neighbor.  The cap $R$ bounds collective rate eigenvalues, including
overlapping channels; its physical angular-rate limit is $RJ_0$, not a
pump-power cap.  We choose $R=8$ as an illustrative budget below the
proposed $\kappa_b=80$--$100$, without identifying it with a hardware limit
or an error threshold for Markov elimination.  The arithmetic band-average noise is
\begin{equation}
\overline{\mathcal N}=\operatorname{Tr}(W_{\mathcal B}\Gamma_g),\qquad
W_{\mathcal B}=\frac1{2\Omega}\int_{-\Omega}^{\Omega}
\frac{yy^\dagger}{\kappa_i|y_i|^2}\,{\mathrm{d}}\omega.
\label{eq:shared_band_objective_prl}
\end{equation}
The matrix $W_{\mathcal B}$ contains the band-averaged response weights.
We denote the minimum band-averaged noise under these constraints
by $\mathcal J^*_{N,R,r}$.
The unrestricted-range problem, denoted by $r=\mathrm{all}$, drops the
range constraint.
Relative to the adaptive pointwise benchmark, uncapped frequency sharing,
the finite rate cap, and restricted range at that cap add nonnegative
noise costs; their decomposition is derived in Sec.~\mbox{S-3\,B} of the
Supplementary Information~\cite{SupplementalMaterial}.

\begin{figure}[htbp]
\centering
\includegraphics[width=0.98\linewidth]{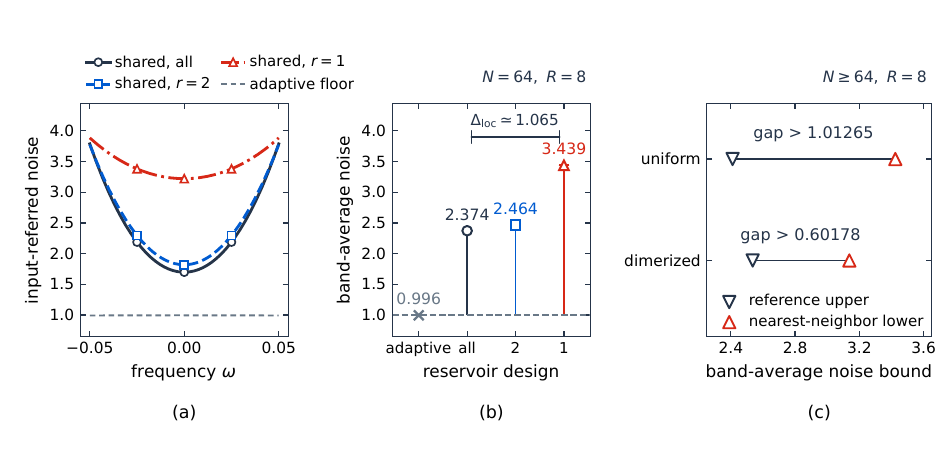}
\caption{Shared-band performance and size-uniform locality cost.
Within each model, shared reservoirs remain fixed across the band at
matched drift and ports, with range $r$ and cap $R$ on the complete rate matrices.
(a,b) Uniform chain, $N=64$, $R=8$, $\Omega=0.05$.
(a) Input-referred noise spectra of certified near-optimal devices:
red triangles, $r=1$; blue squares, $r=2$; black circles, unrestricted range.
The dashed floor permits separate reservoir redesign at each frequency
under the same cap.
(b) Certified minimum band-averaged noises; stems show excess above that floor.
Certification intervals are smaller than markers.
(c) At $R=8$, $\Omega=0.05$, nearest-neighbor lower bounds (red upward
triangles) and achievable reference upper bounds (black downward triangles)
for the uniform and dimerized chains.  Their separation lower-bounds the
locality gap for every finite $N\ge64$.}
\label{fig:shared_cost}
\end{figure}

At $N=64$, $R=8$, and $\Omega=0.05$, the average pointwise floor is
$0.996225$.  Reprogramming the original bond reservoir lowers the
noise from $201.591$ to $3.43857$ at nearest-neighbor range;
range two gives $2.46360$, and unrestricted range $2.37401$.
The locality gap is defined by
\begin{equation}
\Delta_{\rm loc}=\mathcal J^*_{N,R,1}-\mathcal J^*_{N,R,\mathrm{all}}.
\label{eq:locality_gap_definition_nc}
\end{equation}
At this working point, it is approximately $1.06455$
[Fig.~\ref{fig:shared_cost}(a,b)].  Feasible reservoir matrices and
residual-corrected lower bounds enclose both optima.
This objective describes integrated input-referred noise, not the
largest in-band value; the spectra nearly meet at the band edges.
At $R=32,256$, the $N=64$ locality gaps shrink to approximately
$0.0465,0.0431$; thus their magnitude depends strongly on the rate budget.
The adaptive floor permits a different unrestricted reservoir at each
frequency under the same cap.  Finite rates constrain local cancellation
even at one frequency, so the gap combines rate and spatial restrictions.
Methods gives the exact band integration
and bound verification;
additional numerical checks are reported in Secs.~\mbox{S-3\,A,D} of the
Supplementary Information~\cite{SupplementalMaterial}.
Conic solver methods are described in
\cite{DiamondBoyd2016,GoulartChen2024}.
The boundary response below establishes that this separation persists
with length [Fig.~\ref{fig:shared_cost}(c)].

\FloatBarrier

\subsection{Size-uniform locality cost}
The output weights normalized by the input response,
$h_j=y_j/y_1$, obey a left-boundary recurrence.  For the uniform chain,
\begin{equation}
\begin{aligned}
h_1&=1,\qquad h_2=1.3-{\mathrm{i}}\omega,\\
h_{j+1}&=(1.2-{\mathrm{i}}\omega)h_j-\tfrac14h_{j-1}.
\end{aligned}
\label{eq:input_boundary_recurrence_nc}
\end{equation}
The output boundary fixes only the overall response amplitude.
Thus a fixed input-side region carries exactly the same normalized
noise weights at every longer length.  Shared baths can correlate this
region with the tail, so the independent-stage cascade formula
\cite{Friis1944} does not give the locality bound.  Their cross-boundary
contributions must also be bounded at fixed drift.

For a prefix of $m$ input-side sites, its principal rate blocks satisfy
the same local constraints.  The remaining sites form the tail, whose
principal-block noise is nonnegative.
Cross-boundary correlations can lower the total, but positivity and
$0\preceq\Gamma_g\preceq RI$ limit every off-diagonal entry to $R/2$.
A finite-prefix witness together with this boundary correction therefore
gives
\begin{equation}
\mathcal J^*_{N,R,r}(\Omega)\ge
[\alpha_m(\Omega)-R\beta_{m,r}(\Omega)]_+,\qquad N\ge m.
\label{eq:uniform_resource_prl}
\end{equation}
Here $[x]_+=\max(x,0)$.  The prefix witness sets the intercept
$\alpha_m$, while $\beta_{m,r}$ collects the corrections allowed by
the rate cap and correlations across the prefix boundary.
Both coefficients are independent of device length; this is a noise
lower bound for every admissible device.  Methods defines the witness and boundary correction; the full positivity
proof is given in Sec.~\mbox{S-4\,B} of the Supplementary
Information~\cite{SupplementalMaterial}.

An achievable upper bound extends a fixed optimized prefix with local
bond reservoirs, including the bridge bond connecting prefix and tail.
Reserving the bridge rate in the prefix budget
keeps both complete matrices below $8I$ at every length.
The normalized tail weights decay on the certified band, so their
complete noise contribution has a controlled geometric bound.
With $m=64$, the construction has range at most 63, independent of $N$.
Methods gives the extension and its noise ceiling; budget and Floquet
checks are detailed in Secs.~\mbox{S-4\,C,D} of the Supplementary
Information~\cite{SupplementalMaterial}.

For the uniform chain at $\Omega=0.05$, subtracting the feasible
reference ceiling from the nearest-neighbor lower bound yields
\begin{equation}
\mathcal J^*_{N,8,1}-\mathcal J^*_{N,8,\mathrm{all}}>1.01265,
\qquad N\ge64.
\label{eq:certified_locality_gap_prl}
\end{equation}
Thus adding sites cannot eliminate the input-side constraint
[Fig.~\ref{fig:shared_cost}(c)]: reservoir range is a resource independent
of chain length; the gap is certified in Sec.~\mbox{S-4\,F} of the
Supplementary Information~\cite{SupplementalMaterial}.

\FloatBarrier

\subsection{Bandwidth and design targets}
\begin{figure}[htbp]
\centering
\includegraphics[width=\textwidth]{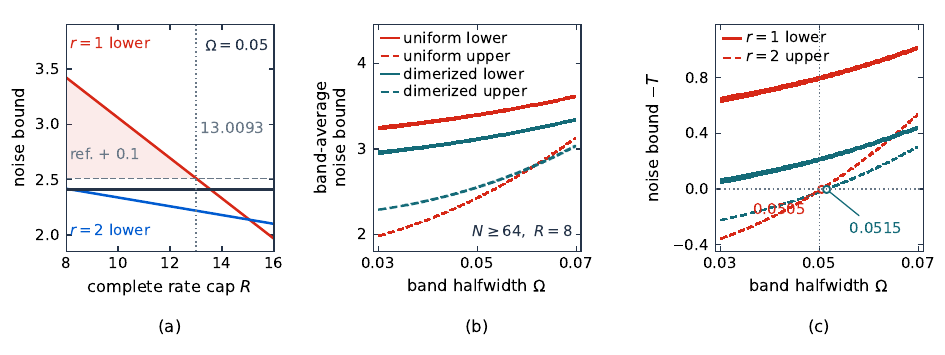}
\caption{Rate and range requirements for fixed broadband reservoirs.
All bounds concern band-averaged input-referred noise at fixed drift and
ports for every finite $N\ge64$.
(a) Uniform chain, $\Omega=0.05$: red and blue, lower bounds for $r=1,2$;
black, the fixed $R=8$ reference ceiling $U_{\rm ref}$;
dashed, the target $U_{\rm ref}+0.1$.
Shading excludes nearest-neighbor attainment; $R>13.0093$ is necessary.
(b,c) $R=8$, with constructions unchanged as $\Omega$ varies.
(b) Nearest-neighbor lower bounds (solid) and reference upper bounds (dashed):
red, uniform; teal, dimerized.
(c) Bounds relative to targets $T=2.6$ (uniform, red) and $T=2.9$
(dimerized, teal): solid, nearest-neighbor lower bound $L_1-T$;
dashed, fixed range-two upper bound $U_2-T$.
Positive $L_1-T$ and negative $U_2-T$ certify $r_{\min}=2$.
Open markers end the certified intervals; the dotted line marks
the design halfwidth $\Omega=0.05$.}
\label{fig:design_targets}
\end{figure}

A dimerized chain alternates the site damping and directional bond couplings,
$(\gamma_1,\gamma_2)=(1.2,1.15)$,
$(t_{R,1},t_{R,2})=(1.15,0.85)$, and
$(t_{L,1},t_{L,2})=(0.15,0.35)$, with the same ports;
the first entry of each pair applies at the input end.
This tests whether the mechanism survives a two-site unit cell.
The Supplementary Information reindexes sites from zero.
Its all-length nearest-neighbor locality gap exceeds $0.60178$ at
$\Omega=0.05$.  Fixing the reference prefix in each family extends the
positive gap to the entire interval $\Omega\in[0.03,0.07]$ at $R=8$:
the uniform and dimerized gaps exceed $0.47876$ and $0.29893$, respectively,
for every finite $N\ge64$ [Fig.~\ref{fig:design_targets}(b)].
The same reservoir couplings are used across all these bandwidths,
with whole-band power gain above $34.97$.

The affine witness also converts a desired noise ceiling into a
necessary rate.  For the uniform chain at $\Omega=0.05$, the fixed $R=8$
reference has certified band-averaged noise ceiling $U_{\rm ref}$.
Reaching $\overline{\mathcal N}\le U_{\rm ref}+0.1$ with nearest-neighbor
reservoirs requires $R>13.0093$
[Fig.~\ref{fig:design_targets}(a)].
Explicit range-two reservoirs attain $\overline{\mathcal N}\le T$,
with $T=2.6$ and $T=2.9$ for the
uniform and dimerized chains, respectively, at $R=8$, $\Omega=0.05$,
while the nearest-neighbor lower bounds exclude both targets.
Keeping both range-two prefixes fixed certifies minimum integer range
exactly two throughout $\Omega\in[0.03,0.0505]$ and $[0.03,0.0515]$,
respectively [Fig.~\ref{fig:design_targets}(c)].
The noise targets are illustrative design choices; the intervals
guarantee attainment without locating physical bandwidth limits.
These rate and range guarantees follow from the bounds in
Secs.~\mbox{S-4\,F,G} of the Supplementary Information~\cite{SupplementalMaterial}.

For noise perturbations at matched response and ports, a gap lower bound
$\Delta_0$ gives a resolvable difference
\begin{equation}
\Delta_{\rm res}\ge\Delta_0-C_Q\delta-2u_{\rm meas}.
\label{eq:noise_tolerance_nc}
\end{equation}
Here $\delta$ bounds the sum of the two force-noise matrix errors in
spectral norm, $u_{\rm meas}$ is the measurement error per device, and $C_Q$ bounds
the band-averaged squared input-normalized response.
The decaying tail makes $C_Q$ independent of length for the two nominal
Markov chains (proof in Sec.~\mbox{S-4\,H} of the Supplementary
Information~\cite{SupplementalMaterial}).  This bound assumes that response matching and
complete-rate feasibility are retained.

\FloatBarrier

\subsection{Resource-calibrated microwave implementation}
The effective rate caps alone do not match the total strength of the
microscopic couplings.  To test this additional resource, theoretical
calculations at $N=64$ retain finite-linewidth auxiliaries,
$10$ kHz intrinsic energy loss and bath occupation $n_b=0.001$, with vacuum ports.
Auxiliary coupling matrices $U_g,U_\ell$ have total squared strength
$\mathcal P_U=\|U_g\|_F^2+\|U_\ell\|_F^2$, determined by the complete trace
$\operatorname{Tr}(\Gamma_g+\Gamma_\ell)$ at fixed auxiliary linewidth and
intrinsic loss (Methods, Eq.~\eqref{eq:m_coupling_resource}).  Matching a cap $\tau$ on this trace as well as $R=8$
tests whether the advantage uses more total coupling strength.
For the uniform chain at the illustrative trace cap
$J_0\tau/(2\pi)=175$ MHz and auxiliary linewidth $80$ MHz, the
nearest-neighbor noise is at least $4.24079$, whereas a fixed range-two
device achieves at most $3.46123$; all 20 tested resource combinations
retain a gap exceeding $0.5719$ quanta.
These 20 nominal comparisons match both spectral and trace caps
[Supplementary Fig.~2(b)].
\par
Calibration errors test a second question: can a device designed after
calibration remain quieter throughout the detection band?
For this parameter grid, only the complete spectral caps are matched.
A range-two reference is redesigned at each calibrated working
point, then held fixed over frequency and halfwidths $45$, $50$, and
$55$ kHz.  It beats a nearest-neighbor benchmark optimized separately
at each halfwidth in all 648 comparisons [Supplementary Fig.~2(a)].
Applying corrections to a nominal reference instead yields positive
certificates in only 274 cases; the remaining comparisons are
uncertified, rather than evidence of a negative optimality gap
(Sec.~\mbox{S-5\,B} of the Supplementary
Information~\cite{SupplementalMaterial}).
\par
To cover continuously varying parameters, interval bounds compare two
feasible uniform-chain seeds.
At each calibrated working point, positive corrections reset their
reservoirs, which are then fixed over frequency and bandwidth.
This continuous comparison matches the complete spectral caps without
a trace cap, and concerns two devices rather than the nearest-neighbor
optimum (Sec.~\mbox{S-5\,C} of the Supplementary
Information~\cite{SupplementalMaterial}).  The certified region covers
independent global drift variations of $\pm0.125\%$, auxiliary linewidths
$80$--$100$ MHz, intrinsic losses $0$--$20$ kHz, occupations $0$--$0.02$,
and halfwidths $30$--$70$ kHz.  Throughout this region, whole-band gain
exceeds $12.23$ dB and the noise difference exceeds $0.48477$ quanta.
A combined force-noise error $\delta J_0/(2\pi)=1$ kHz and $u_{\rm meas}=0.05$
leave a resolvable difference above $0.2927$ quanta.
Complex-response and coupling calibration precede a probe-off heterodyne
noise measurement \cite{LecocqConversion2016,Bernier2017}; division of the
vacuum-subtracted output spectrum by measured $G$ yields $\mathcal N$.
Few-mode experiments \cite{Peterson2017,Bernier2017} and amplifier proposals
\cite{MalzEtAl2018} support component scales.  Independent pump errors,
frequency crowding, compression, and conversion of $\mathcal P_U$ into
pump watts remain to be characterized.

\begin{table}[htbp]
\centering
\small
\begin{tabular}{@{}p{0.22\linewidth}p{0.22\linewidth}p{0.32\linewidth}p{0.14\linewidth}@{}}
\hline
Test & Matched resources & Noise comparison & Gap (quanta) \\
\hline
20 nominal cases & Spectral and total coupling caps &
Nearest-neighbor minimum versus fixed range-two device & $>0.5719$ \\
648 calibrated grid cases & Spectral caps &
Nearest-neighbor minimum at each halfwidth versus fixed range-two device & $>0$ in all cases \\
Continuous calibrated region & Spectral caps &
Two feasible fixed devices & $>0.48477$ \\
\hline
\end{tabular}
\caption{Finite-auxiliary resource comparisons at $N=64$.
Settings may change after calibration but remain fixed over frequency.
The first two tests bound the nearest-neighbor optimum;
the continuous region compares two specified devices.}
\label{tab:resource_comparisons}
\end{table}

\FloatBarrier

\section{Discussion}
In summary, a longer amplifier can increase transmission without
removing the noise constraint imposed by a fixed input-boundary region.
The input-normalized response in that region is unchanged when sites
are added.  A rate-limited nearest-neighbor reservoir therefore cannot
replace the missing spatial correlations by increasing the chain length.
This mechanism distinguishes the all-length locality bound from the
known single-frequency quantum-noise floor and the qualitative
multi-detuning bath-sharing limitation \cite{LauClerk2018}.
\par
The bounds turn this distinction into design criteria for the two
specified nominal Markov chains: insufficient rates exclude noise
targets, while explicit range-two extensions attain selected targets.
The feasible reference used in the unrestricted-range gap bound can have
range up to 63, independently of total length; the separate range-two
constructions establish the minimum-range targets.
At $N=64$, nominal finite-auxiliary tests retain a noise gap under both
spectral and total coupling budgets.  The calibrated grid and continuous
two-device region use spectral caps alone and reset the settings after
calibration.  Their noise-error bounds concern matched responses and
feasible reservoirs, and do not certify an unadjusted device against
independent pump errors or site disorder.
\par
For microwave readout, these results support comparing reservoir
implementations at the same measured transmission and complete rate
budget.  Range two recovers $91.6\%$ of the nearest-neighbor versus
unrestricted noise gap at the chosen $N=64$ working point, giving a
practical intermediate design.  The bounds identify which local
correlations and rate resources a noise target demands, and the
uncertainty needed to resolve the resulting difference.  The proposed
auxiliary-mode factorization does not establish the manufacturability
of a complete 64-mode circuit.  Frequency crowding, unequal auxiliary
linewidths, compression and conversion of coupling budgets into pump
watts require further device characterization.  General range scaling
remains open.  Extensions to nonperiodic bound-state lattices
\cite{WangBoundStates2022}, interacting bosons \cite{li2024non,WangDoublons2026}
or Majorana systems \cite{ZhaoMajorana2021} require separate response and
noise analyses.  Loss-controlled optical and magnomechanical response
\cite{RuanLossEnhanced2025,YuMagnomechanics2025} and dissipative surface-state
interference \cite{DingDissipativeSurface2026} offer further settings in
which to distinguish response enhancement from reservoir fluctuations.

\section{Methods}

\subsection{Microscopic model, dynamics and measured noise}
The signal chain consists of $N$ resonators, with $a_j$ annihilating
a photon at site $j$.  Passive ports address the input site $i=1$ and
output site $o=N$.  Separate damped auxiliary modes mediate correlated
loss and gain on each bond.  In frames rotating at the resonator
frequencies, difference-frequency pumps select photon conversion,
whereas sum-frequency pumps select pair creation.  Resonant pumping
removes the auxiliary detunings in this frame.  The rotating-wave
Hamiltonian of the baseline implementation is
\begin{equation}
\begin{aligned}
\widehat H_{\rm mic}&=\widehat H_{\rm sig}
+\sum_{j=1}^{N-1}(\widehat H_{\ell,j}+\widehat H_{g,j}),\\
\widehat H_{\rm sig}&={\mathrm{i}}\sum_{j=1}^{N-1}v_j^{\rm coh}
(a_{j+1}^\dagger a_j-a_j^\dagger a_{j+1}),\\
\widehat H_{\ell,j}&=g_jb_{\ell,j}^\dagger(a_j-a_{j+1})+\mathrm{h.c.},\\
\widehat H_{g,j}&=g_jb_{g,j}^\dagger(a_j^\dagger+a_{j+1}^\dagger)
+\mathrm{h.c.}.
\end{aligned}
\label{eq:m_microscopic_hamiltonian}
\end{equation}
Here $v_j^{\rm coh}$ is the real coherent hopping rate, $g_j$ the complex
pump-controlled coupling at either end of bond $j$, and $b_{\ell,j}$
and $b_{g,j}$ annihilate independent loss and gain auxiliaries.
The loss interaction exchanges a signal photon with an auxiliary;
the gain interaction creates or annihilates a pair using the pump.
The relative minus sign in the loss coupling is a phase difference,
with all decay rates positive.  Uniform chains have $v_j^{\rm coh}=v$ and $g_j=g$;
alternating couplings implement the dimerized chain.  We set $\hbar=1$.

Each auxiliary loses energy to a vacuum termination at rate $\kappa_b$,
so its amplitude decays at $\kappa_b/2$.  Let $A_\mu$ be the signal
operator coupled to auxiliary $\mu$: $a_j-a_{j+1}$ for loss and
$a_j^\dagger+a_{j+1}^\dagger$ for gain on bond $j$.
The interaction $g_jb_\mu^\dagger A_\mu+\mathrm{h.c.}$ gives
\begin{equation}
\begin{aligned}
\dot b_\mu&=-\frac{\kappa_b}{2}b_\mu-{\mathrm{i}} g_j A_\mu
+\sqrt{\kappa_b}\,b_{{\rm in},\mu},\\
b_{\mu,{\rm sig}}&\simeq-\frac{2{\mathrm{i}} g_j}{\kappa_b}A_\mu,
\qquad L_\mu=\frac{2|g_j|}{\sqrt{\kappa_b}}A_\mu.
\end{aligned}
\label{eq:m_auxiliary_markov}
\end{equation}
Here $b_{{\rm in},\mu}$ is the auxiliary vacuum input field.
The second line applies when the probe detuning $\omega$ and couplings
are small compared with $\kappa_b$, and auxiliary relaxation is faster
than the signal dynamics~\cite{GardinerZoller2004,MetelmannClerk2015}.
Only the signal-induced field is replaced by its adiabatic value;
the vacuum input remains and supplies reservoir fluctuations.
This gives the bond jumps in Eq.~\eqref{eq:bond_jumps_nc}, with
$c_j=4|g_j|^2/\kappa_b$.  The factor four follows from amplitude
decay $\kappa_b/2$ and output coupling $\sqrt{\kappa_b}$.

The reduced signal density operator $\rho$ then obeys~\cite{Lindblad1976,GKS1976}
\begin{equation}
\dot\rho=-{\mathrm{i}}[\widehat H_{\rm sig},\rho]
+\sum_\mu\mathcal D[L_\mu]\rho,\qquad
\mathcal D[L]\rho=L\rho L^\dagger-\tfrac12\{L^\dagger L,\rho\}.
\label{eq:m_effective_master}
\end{equation}
General programmed jumps are $L_\mu^g=\bm a^\dagger g_\mu$ and
$L_\mu^\ell=\ell_\mu^\dagger\bm a$, where
$\bm a=(a_1,\ldots,a_N)^{\mathsf{T}}$ and the columns $g_\mu,\ell_\mu$
contain their complex site amplitudes.  Define
$\Gamma_g=\sum_\mu g_\mu g_\mu^\dagger$ and
$\Gamma_\ell=\sum_\mu\ell_\mu\ell_\mu^\dagger$.
The adjoint dissipator
$\mathcal D^\dagger[L]\bm a=
\{L^\dagger[\bm a,L]+[L^\dagger,\bm a]L\}/2$
gives $+g_\mu g_\mu^\dagger\bm a/2$ for gain and
$-\ell_\mu\ell_\mu^\dagger\bm a/2$ for loss.
Thus the mean amplitudes obey
$\partial_t\langle\bm a\rangle=X\langle\bm a\rangle$, with the drift
in Eq.~\eqref{eq:drift_relation_prl} and
$\widehat H_{\rm sig}=\bm a^\dagger H\bm a$.
For a baseline bond, gain and loss diagonal rates cancel;
their difference contributes $2c_j$ off diagonal.  Coherent hopping
contributes $+v_j^{\rm coh}$ and $-v_j^{\rm coh}$, yielding
$t_{R,j}=c_j+v_j^{\rm coh}$ and $t_{L,j}=c_j-v_j^{\rm coh}$.
On-site energy loss $2\gamma_j$ gives amplitude damping $-\gamma_j$.
The optimized couplings can differ from this baseline while preserving
$\Gamma_g-\Gamma_\ell=M=X+X^\dagger$ and the Hermitian matrix $H$.

The passive port energy linewidths $\kappa_i,\kappa_o$ contribute
$P_\kappa=\kappa_i e_ie_i^\dagger+\kappa_o e_oe_o^\dagger$,
where $e_j$ selects site $j$ and $I$ is the identity.  They are held
separately fixed, with $X_\kappa=X-P_\kappa/2$.
Using a Fourier convention with $\partial_t\to{\mathrm{i}}\omega$, the signal
Langevin equation is
\begin{equation}
({\mathrm{i}}\omega I-X_\kappa)\bm a(\omega)
=-\sqrt{\kappa_i}e_ib_{{\rm in},i}(\omega)
-\sqrt{\kappa_o}e_ob_{{\rm in},o}(\omega)+\bm\xi(\omega).
\label{eq:m_signal_langevin}
\end{equation}
The internal force $\bm\xi=-C_{\ell,{\rm int}}b_{\ell,{\rm in}}
+C_gb_{g,{\rm in}}^\dagger$ contains annihilation inputs for loss
and creation inputs for gain; the columns of $C_{\ell,{\rm int}}$
and $C_g$ are $\ell_\mu$ and $g_\mu$.
Independent vacuum baths give
\begin{equation}
\begin{aligned}
\langle\xi_j^\dagger(t')\xi_k(t)\rangle
&=(\Gamma_g)_{kj}\delta(t-t'),\\
\langle\xi_k(t)\xi_j^\dagger(t')\rangle
&=(\Gamma_\ell)_{kj}\delta(t-t'),\qquad
\langle\xi_k(t)\xi_j(t')\rangle=0.
\end{aligned}
\label{eq:m_force_correlations}
\end{equation}
The gain vacuum supplies spontaneous photons.  Loss vacuum contributes
damping and preserves commutators, but contributes no normally ordered
photon noise above vacuum.

Solving Eq.~\eqref{eq:m_signal_langevin} defines
$\chi_\kappa=({\mathrm{i}}\omega I-X_\kappa)^{-1}$.
The output field is
$b_{{\rm out},o}=b_{{\rm in},o}+\sqrt{\kappa_o}a_o$
\cite{GardinerCollett1985,GardinerZoller2004}.
With $y=\chi_\kappa^\dagger e_o$, its component $y_j^*$ propagates
a fluctuation at site $j$ to the output.  Transmission and noise are
\begin{equation}
\begin{aligned}
S_{oi}&=-\sqrt{\kappa_i\kappa_o}(\chi_\kappa)_{oi},
\qquad G=|S_{oi}|^2=\kappa_i\kappa_o|y_i|^2,\\
S_{\rm out}^{\rm exc}&=\kappa_o y^\dagger\Gamma_g y,
\qquad \mathcal N=\frac{S_{\rm out}^{\rm exc}}{G}.
\end{aligned}
\label{eq:m_output_projection}
\end{equation}
The normally ordered output spectrum follows directly from
Eq.~\eqref{eq:m_force_correlations}; dividing by power gain gives
Eq.~\eqref{eq:input_referred_noise_nc}.  Comparisons assume stable
$X_\kappa$ and nonzero transmission.
The drift depends on the difference of gain and loss, whereas photon
noise depends on gain itself.  Different reservoir couplings can
therefore preserve transmission and reflection while changing noise.

The normally ordered amplifier floor is $1-G^{-1}$
\cite{Caves1982,CavesCombes2012,Clerk2010}.
Indeed, collecting all loss channels, including ports, in $C_\ell$
gives $S=I-C_\ell^\dagger\chi_\kappa C_\ell$ and
$T=C_\ell^\dagger\chi_\kappa C_g$.
The commutator identity $SS^\dagger-TT^\dagger=I$ implies
$(TT^\dagger)_{oo}=\sum_\mu|S_{o\mu}|^2-1\ge G-1$.
Reflection can leave a penalty above this floor even at the fixed-drift
optimum.  For symmetric spectra, symmetrized added noise is
$n_{\rm add}=\mathcal N-(1-G^{-1})/2$; the common offset preserves
matched-response noise differences.

\subsection{Local reservoir interference at a selected frequency}
Vacuum loss baths add no normally ordered photon noise.  Their loss must
be compensated by gain to preserve the drift.  Substituting
$\Gamma_g=M+\Gamma_\ell$ into
Eq.~\eqref{eq:input_referred_noise_nc}, with positivity of both rate matrices,
gives the known bound $\mathcal N\ge f$ in
Eq.~\eqref{eq:pointwise_floor_nc}~\cite{LauClerk2018}.
On the positive branch, the excess is
$y^\dagger\Gamma_\ell y/(\kappa_i|y_i|^2)
=\sum_\mu|\ell_\mu^\dagger y|^2/(\kappa_i|y_i|^2)$.
This is the additional gain noise required by the fixed-drift constraint.
For a bath coupled to two adjacent sites, its two response amplitudes
can cancel at $\omega_0$ by choosing the relative coupling phase and
magnitude.  Each loss channel must therefore be orthogonal to the
measured response.

The adjacent-site vectors $v_j$ in Eq.~\eqref{eq:orthogonal_loss_nc}
implement this condition: $v_j^\dagger y=0$.
Their geometric-mean normalization removes the overall response norm,
retaining the neighboring amplitude ratio and phase.
If every $y_j\ne0$, a vector $x$ satisfying $v_j^\dagger x=0$ on every
bond must have $x_j/y_j$ equal at all sites.
Hence $K=\sum_jv_jv_j^\dagger$ has only $y$
in its kernel.  This output-selected direction need not be a drift eigenmode.

Cancellation alone does not ensure physical compensating gain.
Let $u=y/\|y\|$, $P=I-uu^\dagger$, $s=u^\dagger Mu>0$,
and $m_\perp=PMu$.  The longitudinal gain block is $s$;
$m_\perp$ couples it to transverse directions.
The transverse Schur complement of $M+\alpha K$ is
$PMP+\alpha K|_{y^\perp}-m_\perp m_\perp^\dagger/s$.
Writing $k_0=\lambda_{\min}(K|_{y^\perp})>0$, it is nonnegative if
\begin{equation}
\alpha\ge\frac{\|PMP\|+\|m_\perp\|^2/s}{k_0}.
\label{eq:m_pointwise_schur}
\end{equation}
Thus finite strength makes both $\Gamma_\ell=\alpha K$ and
$\Gamma_g=M+\alpha K$ positive semidefinite while retaining $\Gamma_\ell y=0$.
Matrix and vector norms denote operator and Euclidean norms;
$\|\cdot\|_F$ denotes the Frobenius norm.

A positive semidefinite tridiagonal rate matrix corresponds to actual local baths.
Eliminating its first diagonal pivot subtracts a positive factor
supported on that site and its neighbor; the Schur remainder remains
positive and tridiagonal.  Iteration gives on-site and adjacent-site
jumps.  Zero pivots have zero adjoining entries by positivity;
the block version uses a generalized inverse, as derived in
Sec.~\mbox{S-2\,A} of the Supplementary Information~\cite{SupplementalMaterial}.
Nonzero cell-response blocks allow the same construction for several
modes per cell.  For $s<0$, exchange gain and loss and use $-M$,
giving $(\Gamma_g,\Gamma_\ell)=(\alpha K,-M+\alpha K)$.
At $s=0$, finite attainment requires $Mu=0$; otherwise the floor
is only approached as rates diverge.  A prescribed cap is an additional
restriction, and a different frequency changes the dark response direction.

\subsection{Fixed reservoirs over a finite frequency band}
A frequency scan keeps the same reservoir couplings throughout
$\mathcal B=[-\Omega,\Omega]$.  The measured band average is
\begin{equation}
\mathcal J=\frac1{2\Omega}\int_{-\Omega}^{\Omega}
\mathcal N(\omega)\,{\mathrm{d}}\omega=\operatorname{Tr}(W\Gamma_g),
\qquad W=\langle zz^\dagger\rangle_{\mathcal B},\quad
z=\frac{y}{\sqrt{\kappa_i}|y_i|}.
\label{eq:m_measured_band_average}
\end{equation}
Brackets mean arithmetic band averaging, relevant to integrated
input-referred noise; no worst-case spectral objective is imposed.
The response weight $W$ records
how every spatial noise correlation contributes after input normalization.
The complete matrices obey $0\preceq\Gamma_g,\Gamma_\ell\preceq RI$
and $\Gamma_g-\Gamma_\ell=M$.
The cap $R$ bounds collective rate eigenvalues, including overlapping
channels and on-site loss; external ports remain fixed.
Range $r$ restricts correlations to $|j-k|\le r$.
Banded Cholesky factors couple each bath to at most $r+1$ consecutive sites.
Compensating gain and loss can extend this range while preserving
nearest-neighbor signal drift.

For $R>0$, set $\widetilde L=\Gamma_\ell/R$ and
$\widetilde M=M/R$.  The optimization is
\begin{equation}
\begin{aligned}
p&=\min\operatorname{Tr}(W\widetilde L),\\
\widetilde L,\ \widetilde L+\widetilde M,\ I-\widetilde L,
\ I-\widetilde L-\widetilde M&\succeq0,\\
\mathcal J^*_{N,R,r}&=\operatorname{Tr}(WM)+Rp.
\end{aligned}
\label{eq:m_rate_capped_primal}
\end{equation}
The four constraints impose physical loss, physical gain and their two
complete caps, with $\widetilde L$ restricted to range $r$.
Removing range at the same cap defines the unrestricted comparison.

Exact integration follows from the normalized response
$h_j=y_j/y_1$.  Before reaching the output, the source-free adjoint
equations give the recurrence in Eq.~\eqref{eq:m_boundary_profile},
so $h_j$ is a polynomial of degree $j-1$.
With $x=\omega/\Omega$ and Legendre polynomials $P_\ell(x)$, write
$h_j=\sum_{\ell=0}^{j-1}(-{\mathrm{i}})^\ell c_{j\ell}P_\ell(x)$.
Their orthogonality gives
\begin{equation}
W_{jk}=\frac1{\kappa_i}
\sum_{\ell=0}^{\min(j,k)-1}\frac{c_{j\ell}c_{k\ell}}{2\ell+1}.
\label{eq:m_exact_band_gram}
\end{equation}
The coefficients are rational for rational parameters and halfwidth.
Their triangular matrix has nonzero diagonal for $\Omega>0$,
so $W\succ0$ at every finite length: no nonzero loss matrix is dark
throughout the band.

Without range or rate constraints, let
$A=W^{1/2}MW^{1/2}$ and $F=W^{1/2}\Gamma_gW^{1/2}$.
Then $F,F-A\succeq0$ imply
$\min\operatorname{Tr}F=\operatorname{Tr}A_+$, attained at $F=A_+$.
Here $A_+$ and $A_-=(-A)_+$ are the positive and negative spectral parts.
Projection onto the positive eigenspace of $A$ gives
$\operatorname{Tr}F\ge\operatorname{Tr}A_+$, since $F-A\succeq0$
there and $F\succeq0$ on its complement.
On a positive pointwise branch throughout the band, sharing alone costs
$\operatorname{Tr}A_-$ above $\overline f=\operatorname{Tr}(WM)$.
A negative eigenvalue of $M$ makes this cost positive by congruence.
Nested feasible sets distinguish sharing, cap and locality costs;
finite-length strictness does not yet give a length-independent gap.

Noise comparisons require achievable upper bounds and lower bounds
valid for all allowed reservoirs.  CVXPY and Clarabel provide candidate
matrices~\cite{DiamondBoyd2016,GoulartChen2024}.
For $R>\|M\|$, the interior reservoir
$\widetilde L_0=(I-\widetilde M)/2$ repairs small feasibility violations.
For any real symmetric constraint matrix $T$, a positive shift
$\delta_T$ and factor $F_T$ establish
\begin{equation}
\rho_T=\|T-\delta_TI-F_TF_T^{\mathsf T}\|_\infty<\delta_T
\quad\Longrightarrow\quad T\succeq(\delta_T-\rho_T)I.
\label{eq:m_psd_certificate}
\end{equation}
The maximum absolute row-sum norm bounds the remainder's spectral norm,
so this checks the full rates and their caps.
For the four constraints in Eq.~\eqref{eq:m_rate_capped_primal}, choose
positive matrices $A_d,B_d,C_d,D_d$ and define
$E_d=\Pi_r(W-A_d-B_d+C_d+D_d)$, where $\Pi_r$ retains allowed entries.
The Lagrange identity leaves four nonnegative traces and a residual
$\operatorname{Tr}(E_d\widetilde L)\ge-N\|E_d\|_\infty$, giving
\begin{equation}
p\ge\operatorname{Tr}[(D_d-B_d)\widetilde M]
-\operatorname{Tr}(C_d+D_d)-N\|E_d\|_\infty.
\label{eq:m_rate_capped_dual}
\end{equation}
Multiplying by $R$ and adding $\operatorname{Tr}(WM)$ restores measured
noise units.  Factor recovery, exact rational arithmetic and residual
checks are given in Sec.~\mbox{S-3\,D} of the Supplementary
Information~\cite{SupplementalMaterial}.

\subsection{Boundary contribution to the noise cost}
Dividing the output response by $y_1$ removes its common amplification.
Before the output site, $({\mathrm{i}}\omega I-X_\kappa)^\dagger y=e_o$
has zero source, giving
\begin{equation}
\begin{aligned}
h_1&=1,\qquad h_2=\frac{\gamma_1+\kappa_i/2-{\mathrm{i}}\omega}{t_{R,1}},\\
h_{j+1}&=\frac{(\gamma_j-{\mathrm{i}}\omega)h_j-t_{L,j-1}h_{j-1}}{t_{R,j}}.
\end{aligned}
\label{eq:m_boundary_profile}
\end{equation}
Indices are one-based; 1 and 2 label the entries of each period-one or
period-two parameter pair in Results.  Local damping and backward
hopping fix the next normalized amplitude.  Only the last equation
sets $y_1$, so the first $m$ weights are identical for $N\ge m$.
Because $y(-\omega)=y(\omega)^*$, averaging a feasible reservoir with
its complex conjugate preserves the objective and constraints.
Real reservoirs therefore suffice on the symmetric band.

Correlations with the added chain can cancel part of the prefix noise,
but their size is limited by the complete cap.
For $0\preceq\Gamma_g\preceq RI$, the contraction $2\Gamma_g/R-I$
gives $|(\Gamma_g)_{ij}|\le R/2$ for $i\ne j$.
Each crossing pair therefore contributes at least $-R|W_{ij}|$;
the tail's principal-block noise is nonnegative.  Consequently
\begin{equation}
\mathcal J^*_{N,R,r}\ge\mathcal J^*_{m,R,r}-Rb_{m,r},\qquad
b_{m,r}=\sum_{i\le m<j,\ j-i\le r}|W_{ij}|.
\label{eq:m_prefix_lifting}
\end{equation}
There are finitely many crossing pairs at fixed range, independently
of chain length.  Including absent pairs in shorter tails enlarges
the correction conservatively.

Write the prefix rates as $G_m=L_m+M_m$, distinct from power gain $G_N$.
Positive matrices $A,B,C,D$ multiply its four positivity and cap
constraints.  Define
\begin{equation}
\begin{aligned}
E&=\Pi_r(W_m-A-B+C+D),\\
\Lambda(E)&=\sum_i(-E_{ii})_++\sum_{i<j}|E_{ij}|,\\
\alpha_m&=\operatorname{Tr}(W_mM_m)+\operatorname{Tr}[(D-B)M_m],\\
\beta_{m,r}&=\operatorname{Tr}(C+D)+\Lambda(E)+b_{m,r},\\
\mathcal J^*_{N,R,r}&\ge\alpha_m-R\beta_{m,r}.
\end{aligned}
\label{eq:m_prefix_witness}
\end{equation}
The identity behind this bound expresses
$\operatorname{Tr}(W_mG_m)$ as
$\alpha_m-R\operatorname{Tr}(C+D)+\operatorname{Tr}(EL_m)$
plus four nonnegative constraint traces.
Diagonal entries of $L_m$ lie in $[0,R]$ and off-diagonal entries
have modulus at most $R/2$, so the residual is at least
$-R\Lambda(E)$.  Subtracting the crossing correction proves the bound;
the full identity is given in Sec.~\mbox{S-4\,B} of the Supplementary
Information~\cite{SupplementalMaterial}.
These constants depend on bandwidth and range, but not $N$.
For $\beta_{m,r}>0$, noise target $T$ requires
$R\ge[\alpha_m-T]_+/\beta_{m,r}$.
Fixed factors remain valid at changed budgets; this gives a necessary
rate, not a sufficient rate or universal optimal scaling law.

\subsection{Reservoir extension and bandwidth-dependent design targets}
To turn the lower bound into a locality cost, a quieter comparison
reservoir must remain feasible at every longer chain.
Start with a prefix pair of difference $M_m$ and reserve the rate
$b_m$ for the bridge bond joining the prefix and tail, where
$b_j=t_{R,j}+t_{L,j}$.
Append bond pairs
$b_j(e_j\pm e_{j+1})(e_j\pm e_{j+1})^{\mathsf T}/2$
for $j=m,\ldots,N-1$, with $+$ for gain and $-$ for loss,
and on-site loss $2\gamma_je_je_j^{\mathsf T}$ for $j>m$.
Their diagonal bond contributions cancel in the difference,
and their off-diagonal contributions restore $M_N$.
For period at most two, sufficient complete caps are
\begin{equation}
G_m,L_m\preceq(R_0-b_m)I,\qquad
2\gamma_{\max}+b_1+b_2\le R_0-b_m.
\label{eq:m_extension_budget}
\end{equation}
The prefix--tail direct sum is bounded by its larger block norm;
adding the bridge costs at most $b_m$ per matrix.
At $m=64$, $R_0=8$, prefix caps are $6.75$ and $6.8$ for the uniform
and dimerized chains, with bridge norms $1.25$ and $1.2$ and tail loss
norms at most $4.9$.  The bridge is omitted at $N=m$.
Odd lengths are included.  An unrestricted prefix needs range at most
63 independently of $N$; a range-two prefix retains range two.

The added sites' noise is controlled by their input-normalized response.
The two-site transfer roots solve $\lambda^2-\tau_F\lambda+d=0$,
where $d=t_{L,1}t_{L,2}/(t_{R,1}t_{R,2})$ and
$\tau_F=(\gamma_1-{\mathrm{i}}\omega)(\gamma_2-{\mathrm{i}}\omega)/(t_{R,1}t_{R,2})
-t_{L,1}/t_{R,2}-t_{L,2}/t_{R,1}$.
These are spatial roots, not temporal stability eigenvalues.
Choose rational $\sqrt d<q<1$, bounds
$|\operatorname{Re}\tau_F|\le T_F$, $|\operatorname{Im}\tau_F|\le B_F$
and $\operatorname{Re}\tau_F\ge a_F>0$, such that
\begin{equation}
\frac{T_F^2}{(q+d/q)^2}+\frac{B_F^2}{(q-d/q)^2}<1,\qquad
a_F^2-B_F^2-4d\ge\sigma^2>0.
\label{eq:m_floquet_conditions}
\end{equation}
The first condition keeps the transfer trace inside the ellipse traced
by $\lambda+d/\lambda$ on $|\lambda|=q$, bounding both roots by $q$.
The second separates the roots by $\sigma$ and bounds the smaller modulus
by $s_F=2d/(a_F+\sigma)<1$, controlling the response prefactors.
For parity $p=1,2$, bounds $H_p\ge|h_p|$ and $H_{p+2}\ge|h_{p+2}|$
give amplitudes $P_p=(H_{p+2}+s_FH_p)/\sigma$ and
$Q_p=(H_{p+2}+qH_p)/\sigma$.
For integers $k,K\ge0$, the squared envelope sums to
\begin{equation}
\begin{aligned}
|h_{2k+p}|&\le P_pq^k+Q_ps_F^k,\\
\sum_{j\ge2K+1}\frac{|h_j|^2}{\kappa_i}&\le\mathcal T_K:=
\frac1{\kappa_i}\sum_{p=1}^{2}\left[
\frac{P_p^2q^{2K}}{1-q^2}+\frac{2P_pQ_p(qs_F)^K}{1-qs_F}
+\frac{Q_p^2s_F^{2K}}{1-s_F^2}\right].
\end{aligned}
\label{eq:m_floquet_tail}
\end{equation}

With a site cutoff $j_{\rm cut}=160$, the prefix, explicit added bonds
and remaining geometric tail give
\begin{equation}
\begin{aligned}
U_m(\Omega)={}&\operatorname{Tr}(W_mG_m)
+\sum_{j=m}^{j_{\rm cut}-2}\frac{b_j}{2}
(W_{jj}+2\operatorname{Re}W_{j,j+1}+W_{j+1,j+1})\\
&+(b_1+b_2)\mathcal T_{(j_{\rm cut}-2)/2}.
\end{aligned}
\label{eq:m_extension_noise}
\end{equation}
Shorter chains omit positive bonds; longer chains obey the same tail bound.

Stability is proved separately: the margin
$\gamma_{\min}-\sqrt{t_{R,1}t_{L,1}}-\sqrt{t_{R,2}t_{L,2}}>0$
makes $X_\kappa$ similar to a negative-definite symmetric Jacobi matrix.
If $a_n>0$ are the eigenvalues of $-X_\kappa$, then
$G_N(\omega)=\kappa_i\kappa_o(\prod_{j=1}^{N-1}t_{R,j})^2/
\prod_{n=1}^{N}(a_n^2+\omega^2)$, minimized at band endpoints.
Endpoint checks and tail bounds cover finite lengths without assuming
an infinite-chain steady state.

For a fixed reservoir, nonnegative noise gives
$\mathcal J_{\Gamma_g}(\Omega)\le(b/a)\mathcal J_{\Gamma_g}(b)$
for $\Omega\in[a,b]$.  A narrower band $s\le a$ is contained in every
such band, so any feasible local reservoir also obeys
$\mathcal J_{\Gamma_g}(\Omega)\ge(s/b)\mathcal J^*_{N,R,1}(s)$.
Combining these integral inclusions with the prefix bound gives
\begin{equation}
L_{[a,b]}(R)=\max_{s\in\mathcal S,\ s\le a}\frac{s}{b}
[\alpha_m(s)-R\beta_{m,1}(s)]_+,\qquad
U^{\rm ref}_{[a,b]}=\frac{b}{a}U_m^{\rm ref}(b),\quad
U^2_{[a,b]}=\frac{b}{a}U_m^2(b).
\label{eq:m_bandwidth_enclosure}
\end{equation}
The superscripts distinguish the reference and range-two constructions.
Saved witnesses cover every halfwidth in $[0.03,0.07]$.
$L>U^{\rm ref}$ proves locality cost; $U^2<T<L$ excludes nearest-neighbor
attainment while providing a range-two device meeting the target.
The all-finite-$N\ge64$ noise guarantees apply to the two specified
nominal Markov chains.  Budget and root checks are detailed in
Secs.~\mbox{S-4\,C--E} of the Supplementary Information~\cite{SupplementalMaterial}.

\subsection{Finite reservoir memory and coupling resources}
The same auxiliary model can retain memory by keeping the frequency
denominator in Eq.~\eqref{eq:m_auxiliary_markov} instead of its adiabatic
limit.  Program one auxiliary per jump using
$\widehat H_{\rm int}=\sum_\mu[b_{\ell,\mu}^\dagger
u_{\ell,\mu}^\dagger\bm a+b_{g,\mu}^\dagger\bm a^\dagger u_{g,\mu}
+\mathrm{h.c.}]$, where $u_{\ell,\mu},u_{g,\mu}$ are physical coupling columns.
This conjugation convention gives effective jump amplitudes
$\ell_\mu=2u_{\ell,\mu}/\sqrt{\kappa_b}$ and
$g_\mu=2u_{g,\mu}/\sqrt{\kappa_b}$ up to channel phases.
The matrices $U_\ell,U_g$ collect these columns.
Include intrinsic energy loss $D_{\rm int}=\eta I$ and define
$C_{\rm coh}=(X-X^\dagger)/2=-{\mathrm{i}} H$.
Programmed loss must satisfy $\Gamma_\ell-D_{\rm int}\succeq0$;
full-rank intrinsic loss prevents exact positive-branch Markov saturation.

For common resonant linewidths, exact elimination within the linear
rotating-wave model gives
\begin{equation}
X_{\rm eff}=C_{\rm coh}-\frac{D_{\rm int}+P_\kappa}{2}
+\frac{\alpha(\omega)}2(M+D_{\rm int}),\qquad
\alpha(\omega)=(1+2{\mathrm{i}}\omega/\kappa_b)^{-1}.
\label{eq:m_finite_auxiliary_response}
\end{equation}
Here $\alpha(\omega)$ is the dimensionless auxiliary response filter,
distinct from the pointwise loss strength $\alpha$.
At zero detuning $\alpha(0)=1$, recovering $X_\kappa$.
Common coherent and net drifts, ports, intrinsic loss and linewidths
therefore preserve the full response for different reservoir decompositions.
The finite filter changes magnitude and phase; the signal subsystem
alone no longer obeys the Markov master equation above.
For common auxiliary and intrinsic occupation $n_b$ and vacuum ports,
the thermal correlations yield
\begin{equation}
Q(\omega)=|\alpha|^2[(n_b+1)\Gamma_g
+n_b(\Gamma_\ell-D_{\rm int})]+n_bD_{\rm int}.
\label{eq:m_finite_auxiliary_noise}
\end{equation}
The three terms describe stimulated gain, programmed thermal loss and
intrinsic thermal loss.  With
$\chi=({\mathrm{i}}\omega I-X_{\rm eff})^{-1}$ and its input-normalized output
response $z$, noise is $z^\dagger Qz$.
Substituting $\Gamma_\ell=\Gamma_g-M$ and comparing matched devices gives
\begin{equation}
\Delta_{12}=\frac{1+2n_b}{2\Omega}\int_{-\Omega}^{\Omega}
|\alpha|^2z^\dagger(\Gamma_{g,1}-\Gamma_{g,2})z\,{\mathrm{d}}\omega.
\label{eq:m_finite_device_gap}
\end{equation}
Here $\Delta_{12}=\overline{\mathcal N}_1-\overline{\mathcal N}_2$.
Common offsets cancel, although absolute noise rises with occupation.
Optimization uses $V=(1+2n_b)\langle|\alpha|^2zz^\dagger\rangle$
and offset $q_{\rm th}=n_b\langle z^\dagger[D_{\rm int}
-|\alpha|^2(M+D_{\rm int})]z\rangle$, giving
$\operatorname{Tr}(V\Gamma_g)+q_{\rm th}$.

The total squared coupling strength is
\begin{equation}
\begin{aligned}
\Gamma_g&=\frac4{\kappa_b}U_gU_g^\dagger,\qquad
\Gamma_\ell=D_{\rm int}+\frac4{\kappa_b}U_\ell U_\ell^\dagger,\\
\mathcal P_U&=\|U_g\|_F^2+\|U_\ell\|_F^2
=\frac{\kappa_b}{4}[\operatorname{Tr}(\Gamma_g+\Gamma_\ell)-N\eta].
\end{aligned}
\label{eq:m_coupling_resource}
\end{equation}
At fixed linewidth and intrinsic loss, a trace cap
$\operatorname{Tr}(\Gamma_g+\Gamma_\ell)\le\tau$ matches this resource.
Unlike $R$, it bounds total rather than largest collective rate.
Model rates multiply by $J_0$ to give angular rates; reported frequencies
divide by $2\pi$.  Physical coupling strength is $J_0^2\mathcal P_U$,
not pump watts, which additionally require mixer and line calibration.

The optimization in Eq.~\eqref{eq:m_rate_capped_primal} now requires
$\widetilde L-D_{\rm int}/R\succeq0$ and
$2\operatorname{Tr}\widetilde L\le t=(\tau-\operatorname{Tr}M)/R$.
Use $V$ instead of $W$ and a trace multiplier $\xi_{\rm tr}\ge0$.
Adding $2\xi_{\rm tr}I$ to $E_d$ and
$\operatorname{Tr}(A_dD_{\rm int}/R)-\xi_{\rm tr}t$ to
Eq.~\eqref{eq:m_rate_capped_dual} gives the physical lower bound.
Multiply by $R$ and restore $\operatorname{Tr}(VM)+q_{\rm th}$.
Upper bounds use an interior reservoir with both spectral and trace
slack.  Programmed-loss variables and the saved factors' thermal
rescaling are detailed in Sec.~\mbox{S-5\,B} of the Supplementary
Information~\cite{SupplementalMaterial}.

\subsection{Calibration, robustness and noise resolution}
Calibration first determines complex transmission and reflection,
couplings and intrinsic loss.  Reservoir settings are chosen for each
calibrated working point and then kept fixed during frequency and
bandwidth scans.  With the coherent tone off, pumps and ports remain
on and the input is terminated in a calibrated cold load.
A reference-temperature sweep determines receiver gain, loss, noise
and the heterodyne vacuum offset~\cite{LecocqConversion2016,Bernier2017}.
Subtract this background, divide by measured power gain and average
over identical bands.

Three calculations test distinct constraints at $N=64$.
The 20 nominal comparisons match $R=8$ and total trace caps
$\tau\in\{175,190,210,250,325\}$ for both ranges, with
$\eta=0.01$, $n_b=0.001$, $\Omega=0.05$ and
$\kappa_b\in\{80,100\}$ for both drifts.
Nearest-neighbor lower bounds minus range-two feasible upper bounds
exceed $0.57191$.

The calibrated grid instead matches spectral caps alone: independent
global changes $-0.005,0,0.005$ in damping and both directed hoppings,
two linewidths and $\eta\in\{0,0.02\}$ give 216 devices.
A range-two reservoir designed at $\Omega=0.05$, cap $7.8$, is frozen
across halfwidths $0.045,0.05,0.055$.
All 648 comparisons beat independently optimized nearest-neighbor
benchmarks at cap $8$, $n_b=0.001$; Eq.~\eqref{eq:m_finite_device_gap}
extends their signs to $n_b\in[0,0.02]$.
Corrected nominal references certify only 274 cases, motivating
calibration before design; failed bounds do not imply reversed optima.

The continuous test compares two feasible uniform-chain devices of
range one and two, designed at $\Omega=0.05$, cap $7.8$.
Common positive on-site and bond corrections realize calibrated
$\Delta M$, with $\eta I$ added to both rates.
For a bond change $d=\Delta M_{j,j+1}$, vectors
$e_j\pm\operatorname{sgn}(d)e_{j+1}$ have weight $|d|/2$;
their gain-minus-loss difference supplies that bond change.
Coherent conversion is also recalibrated.
These common corrections preserve
$\Delta G=\Gamma_{g,1}-\Gamma_{g,2}$ and keep the complete cap below
$7.8445$.  Settings vary between working points but remain fixed
within each scan.  No trace cap or global nearest-neighbor optimum
is asserted.  The broad region has drift changes $\pm0.5\%$,
$\kappa_b\in[80,100]$, $\eta,n_b\in[0,0.02]$ and
$\Omega\in[0.03,0.07]$.

Continuous coverage requires bounding response between sampled points.
Let $h_j=(\chi^\dagger e_o)_j/(\chi^\dagger e_o)_i$.
Intervals enclose recurrence coefficients, with central response
$\widehat h$.  Subtracting its recurrence turns coefficient changes
into forcing; the central spatial Green function bounds
$|h_j-\widehat h_j|\le\varepsilon_j$.
Expanding the noise projection gives
\begin{equation}
\left|h^\dagger\Delta Gh-\widehat h^\dagger\Delta G\widehat h\right|
\le2\sum_j|(\Delta G\widehat h)_j|\varepsilon_j
+\sum_{jk}|\Delta G_{jk}|\varepsilon_j\varepsilon_k.
\label{eq:m_interval_noise}
\end{equation}
Matrix uncertainty $E^G_{jk}\ge|\delta\Delta G_{jk}|$ adds
$\sum_{jk}E^G_{jk}(|\widehat h_j|+\varepsilon_j)
(|\widehat h_k|+\varepsilon_k)$.
Outward rounding and roundoff allowances cover complete and partial
frequency cells after multiplying by $|\alpha|^2/\kappa_i$ and
$1+2n_b$.  The broad region gives gain above $3.335778$ and separation
above $0.419615$; changes $\pm0.125\%$ give gain above $16.745077$
and separation above $0.484771$.

For matched responses, ports and feasible reservoirs, let combined
force-noise uncertainty obey
$\sup_{\mathcal B}(\|\Delta Q_1\|+\|\Delta Q_2\|)\le\delta$
and $C_Q\ge\langle\|z\|^2\rangle$.  The quadratic-form bound gives
\begin{equation}
\begin{aligned}
|\delta\overline{\mathcal N}_r|&\le\frac1{2\Omega}
\int_{-\Omega}^{\Omega}\|\Delta Q_r\|\,\|z\|^2\,{\mathrm{d}}\omega,\\
\Delta_{\rm res}&\ge\Delta_0-C_Q\delta-2u_{\rm meas}.
\end{aligned}
\label{eq:m_noise_uncertainty}
\end{equation}
Here $\Delta_0$ is unperturbed separation and $u_{\rm meas}$ bounds
each device's measurement error.  Broad and high-gain regions have
$C_Q<115.258221$ and $C_Q<92.060749$; at $\delta=0.001$,
$u_{\rm meas}=0.05$, resolvable gaps exceed $0.204357$ and $0.292710$
quanta.  For the two nominal Markov chains, boundary compatibility
and the tail bound also control $C_Q=\operatorname{Tr}W_N$
for every finite $N\ge64$.  Finite-auxiliary noise comparisons remain
at $N=64$.

Finite-auxiliary stability is checked independently of spatial decay.
Choose a diagonal similarity $S_{\rm sim}$ such that
$A=S_{\rm sim}^{-1}X_\kappa S_{\rm sim}$ is symmetric and negative
definite, and set $x=S_{\rm sim}^{-1}\langle\bm a\rangle$ and
$B=S_{\rm sim}^{-1}(M+D_{\rm int})S_{\rm sim}/2$.
Let $\beta_\ell=(\langle b_{\ell,\mu}\rangle)_\mu$ and
$\beta_g=(\langle b_{g,\mu}^\dagger\rangle)_\mu$ collect auxiliary means.
Their feedback on the signal is
$u_{\rm fb}=-{\mathrm{i}} S_{\rm sim}^{-1}(U_\ell\beta_\ell+U_g\beta_g)$.
The auxiliary equations give
$\dot u_{\rm fb}=-\kappa_bu_{\rm fb}/2+\kappa_b Bx/2$, while
$\dot x=(A-B)x+u_{\rm fb}$.
The memory correction $\zeta=u_{\rm fb}-Bx$ therefore obeys
\begin{equation}
\dot x=Ax+\zeta,\qquad
\dot\zeta=-BAx-(\kappa_bI/2+B)\zeta.
\label{eq:m_memory_dynamics}
\end{equation}
For $-A\succeq\mu I$, $\|A\|\le a$, $\|B\|\le b$,
the energy $\|x\|^2+\|\zeta\|^2/(ba)$ decays if
$\mu(\kappa_b/2-b)>ba$; the broad box satisfies this for all finite
$N\ge64$.  Auxiliary combinations absent from $u_{\rm fb}$ decay at
$\kappa_b/2$.  The energy estimate, interval implementation and independent
checks are given in Secs.~\mbox{S-5\,B,C} and S-6 of the Supplementary
Information~\cite{SupplementalMaterial}.
Unmatched linewidths, independent pump errors, disorder and compression
require separate characterization.

\section{Data availability}
The numerical data, reservoir matrices and certificates supporting
the main and supplementary results are public in repository version
1.3.1~\cite{LiReservoirCode2026}, including calibrated-device comparisons,
parameter-region bounds, trace-budget tests and noise-error certificates.
Core data are also supplied as Supplementary Data; the full reservoir
and matrix-witness banks are provided in documented public archives.
An English guide maps numerical claims to saved results and gives
file-level checksums and validation commands.
No experimental data were collected.

\section{Code availability}
The Python code, parameters, solver settings and reproduction commands
for the baseline, calibrated-device and noise-tolerance calculations
are public in repository version 1.3.1~\cite{LiReservoirCode2026}.
Core code and inputs are also supplied as Supplementary Data.
The English guide records dependencies and distinguishes saved-result
verification from optimizer reruns, with the scope and expected outputs
of each check.

\begin{samepage}
\section*{Acknowledgements}
This work was supported by the National Key Research and Development
Program of China (Grant Nos. 2021YFA0718300, 2021YFA1400900, and
2024YFF0726700), the National Natural Science Foundation of China
(Grant Nos. 12334012, 12234012, and 52327808), the Space Application
System of China Manned Space Program, and the Elite Revitalizing
Inner Mongolia Program (Grant No. 2025TGL05).
\par
\end{samepage}

\section*{Author contributions}
Chengxi Li performed the theoretical calculations and numerical simulations
and carried out the majority of the research. Wanzi Sun contributed research
ideas. Wuming Liu provided research support. Hao Zhu designed the proposed
experimental setup. Tie-Fu Zhang and Lin Zhuang contributed to discussions
of the results.

\section*{Competing interests}
The authors declare no competing interests.

\end{document}

\begin{center}
{\Large\bfseries Supplementary Information\par}
\vspace{0.5em}
{\large\bfseries{The noise cost of reservoir locality in broadband quantum amplifier chains}\par}
\vspace{0.7em}
Chengxi Li$^{1,2}$, Hao Zhu$^{1,2}$, Tie-Fu Zhang$^{2,3}$, Lin Zhuang$^4$,
Wanzi Sun$^{2,5,*}$ and Wuming Liu$^{2,*}$\par
\vspace{0.7em}
\end{center}
{\small
$^1${Beijing Institute of Petrochemical Technology, Beijing 102617, China}\par
$^2${Beijing National Laboratory for Condensed Matter Physics, Institute of Physics, Chinese Academy of Sciences, Beijing 100190, China}\par
$^3${Faculty of Science, Kunming University of Science and Technology, Kunming 650500, China}\par
$^4${State Key Laboratory of Optoelectronic Materials and Technologies, School of Physics, Sun Yat-Sen University, Guangzhou, 510275, China.}\par
$^5${Jining University, Qufu, Shandong 273155, China}\par
\medskip
$^*$Correspondence: Wanzi Sun
(\href{mailto:wzsun@jnxy.edu.cn}{wzsun@jnxy.edu.cn}) and Wuming Liu
(\href{mailto:wmliu@iphy.ac.cn}{wmliu@iphy.ac.cn}).\par}

Sections \ref{sec:s_model}--\ref{sec:s_size_uniform} give extended
proofs and certificate details for the calculations in the main Methods.
Section \ref{sec:s_microwave} supplies implementation parameters and
additional tests; Sec.~\ref{sec:s_verification} records verification
and data access.  All-length guarantees concern the two nominal Markov
chains; finite-auxiliary noise comparisons are at $N=64$.

\tableofcontents
\clearpage

\section*{Supplementary Methods}

\section{Model and measured noise}
\label{sec:s_model}

The signal modes obey the main-text Lindblad equation
\cite{Lindblad1976,GKS1976}, with
$\widehat H=\bm a^\dagger H\bm a$ and $H=H^\dagger$.
Linear channels $L_\mu^g=\bm a^\dagger g_\mu$ and
$L_\mu^\ell=\ell_\mu^\dagger\bm a$ give
${\mathrm{d}}\langle\bm a\rangle/{\mathrm{d}} t=X\langle\bm a\rangle$, with
$X=-{\mathrm{i}} H+(\Gamma_g-\Gamma_\ell)/2$
\cite{ProsenSeligman2010,BarthelZhang2022}.
Here $\Gamma_g=\sum_\mu g_\mu g_\mu^\dagger$ and
$\Gamma_\ell=\sum_\mu\ell_\mu\ell_\mu^\dagger$ are positive semidefinite
(PSD) rate matrices.

\subsection{Local realization and stability}
For $t_R,t_L,\gamma\ge0$, set $c=(t_R+t_L)/2$ and $v=(t_R-t_L)/2$.
Choose Hermitian hopping $H_{j+1,j}={\mathrm{i}} v$, $H_{j,j+1}=-{\mathrm{i}} v$ and
the bond jumps
\begin{equation}
L^g_j=\sqrt c(a_j^\dagger+a_{j+1}^\dagger),\qquad
L^\ell_j=\sqrt c(a_j-a_{j+1}).
\label{eq:s_bond_jumps}
\end{equation}
with onsite loss $L_j^0=\sqrt{2\gamma}\,a_j$.
The bond vectors $g_j=\sqrt c(e_j+e_{j+1})$ and
$\ell_j=\sqrt c(e_j-e_{j+1})$ have difference
$g_jg_j^\dagger-\ell_j\ell_j^\dagger
=2c(e_je_{j+1}^{\mathsf{T}}+e_{j+1}e_j^{\mathsf{T}})$.
Combining this with $-{\mathrm{i}} H$ gives $X_{jj}=-\gamma$,
$X_{j+1,j}=c+v=t_R$, and $X_{j,j+1}=c-v=t_L$.
The matrices $\Gamma_g=\sum_jg_jg_j^\dagger$ and
$\Gamma_\ell=2\gamma I+\sum_j\ell_j\ell_j^\dagger$ are PSD, establishing
complete positivity (CP) of this local realization.
For $t_R>t_L>0$, the open-chain drift is diagonally similar to a Hermitian
matrix, with eigenvalues
$\lambda_m=-\gamma+2\sqrt{t_Rt_L}\cos[m\pi/(N+1)]$.
Thus finite-chain stability requires
$\gamma>2\sqrt{t_Rt_L}\cos[\pi/(N+1)]$; a nonzero limiting damping gap
requires $\gamma>2\sqrt{t_Rt_L}$.  Finite- and infinite-chain stability can
differ \cite{UghrelidzeEtAl2024}.  The unprobed gap tends to $0.2$ at the
working point, but need not determine non-normal transient response
\cite{TrefethenEmbree2005,Haga2021,MoriShirai2020,MoriShirai2023}.

For the uniform chain, define $s=\sqrt{t_Rt_L}$ and
$q=\gamma/(2s)+\sqrt{[\gamma/(2s)]^2-1}$.  At zero detuning, the endpoint
response exponent is $\Lambda_{\rm resp}=\log[\sqrt{t_R/t_L}/q]$.
It is positive in the stable amplification window
$2s<\gamma<t_R+t_L$.  This standard response property explains the
working point \cite{XueDirectional2020,LiWan2021,LiuStable2023};
the reservoir comparisons below retain the exact
finite-size port damping.

\subsection{Response and input-output detection}
\label{sec:s_observables}

Before port loading, $\chi(\omega)=({\mathrm{i}}\omega I-X)^{-1}$ gives endpoint
response $\chi_{N1}=e_N^{\mathsf{T}}\chi e_1$ and noise
$S_N=e_N^{\mathsf{T}}\chi\Gamma_g\chi^\dagger e_N$.
For stable $X$, integrating $S_N$ with measure ${\mathrm{d}}\omega/(2\pi)$ gives
the steady endpoint occupation.  The following optimization instead
uses input-referred noise with fixed detection ports.

Passive ports $L_{\rm in}=\sqrt{\kappa_{\rm in}}\,a_1$ and
$L_{\rm out}=\sqrt{\kappa_{\rm out}}\,a_N$ have energy rates
$\kappa_{\rm in}=\kappa_i$ and $\kappa_{\rm out}=\kappa_o$.
They give the exact drift
$X_\kappa=X-\kappa_{\rm in}e_1e_1^{\mathsf{T}}/2
-\kappa_{\rm out}e_Ne_N^{\mathsf{T}}/2$ and Langevin equation
\begin{equation}
({\mathrm{i}}\omega I-X_\kappa)a(\omega)
=-\sqrt{\kappa_{\rm in}}\,e_1 b_{{\rm in},1}(\omega)
-\sqrt{\kappa_{\rm out}}\,e_N b_{{\rm in},N}(\omega)
+F(\omega),
\label{eq:s_langevin_io}
\end{equation}
where $a=(a_1,\ldots,a_N)^{\mathsf{T}}$.
The input-output relation
$b_{{\rm out},N}=b_{{\rm in},N}+\sqrt{\kappa_{\rm out}}\,a_N$
\cite{GardinerCollett1985,GardinerZoller2004} gives the transmission
\begin{equation}
S_{N1}(\omega)=-\sqrt{\kappa_{\rm in}\kappa_{\rm out}}\,
[\chi_\kappa(\omega)]_{N1},\qquad
G(\omega)=|S_{N1}(\omega)|^2,
\label{eq:s_transmission}
\end{equation}
where $\chi_\kappa=({\mathrm{i}}\omega I-X_\kappa)^{-1}$ includes port damping
exactly in every optimization and certificate.

For noise detection, the internal force contains both loss and gain inputs,
$F=-C_{\ell,{\rm int}}b_{\ell,{\rm in}}+C_gb_{g,{\rm in}}^\dagger$.
Independent vacuum inputs give
$\langle F_j^\dagger(t')F_i(t)\rangle=(\Gamma_g)_{ij}\delta(t-t')$,
$\langle F_i(t)F_j^\dagger(t')\rangle=(\Gamma_\ell)_{ij}\delta(t-t')$,
and $\langle F_i(t)F_j(t')\rangle=0$.
Only the creation inputs contribute to the normally ordered output photon
noise above vacuum.  Its spectrum is
\begin{equation}
S_{{\rm out},N}^{\rm exc}(\omega)
=\kappa_{\rm out}\,
e_N^{\mathsf{T}}\chi_\kappa(\omega)\Gamma_g
\chi_\kappa^\dagger(\omega)e_N,
\label{eq:s_output_noise}
\end{equation}
In the noninvasive-probe limit this reduces to
$S_{{\rm out},N}^{\rm exc}(\omega)=\kappa_{\rm out}S_N(\omega)$.
The homodyne quadrature
$X_\theta^{\rm out}(\omega)=e^{-{\mathrm{i}}\theta}b_{{\rm out},N}(\omega)
+e^{{\mathrm{i}}\theta}b_{{\rm out},N}^\dagger(-\omega)$ has symmetrized spectrum
\begin{equation}
\begin{aligned}
S_{X_\theta}^{\rm out}(\omega)
&=1+S_{{\rm out},N}^{\rm exc}(\omega)+S_{{\rm out},N}^{\rm exc}(-\omega)\\
&\longrightarrow1+\kappa_{\rm out}[S_N(\omega)+S_N(-\omega)]
\end{aligned}
\label{eq:s_homodyne}
\end{equation}
for this phase-preserving block, with the vacuum quadrature floor normalized
to unity; the second line is the noninvasive-probe limit.  For a spectrum
symmetric under $\omega\to-\omega$, this becomes $1+2S_{\rm out}^{\rm exc}$.
The symmetrized added noise referred to a vacuum input is therefore
$n_{\rm add}=\mathcal N_{\rm in}^{\rm exc}-(1-G^{-1})/2$.
Its quantum limit is $(1-G^{-1})/2$, whereas the normally ordered convention
used in the optimization has floor $1-G^{-1}$ for $G>1$.
These conventions give the same fixed-drift noise differences
\cite{HausMullen1962,Caves1982,CavesCombes2012,Clerk2010}.
Phase-sensitive
anomalous spectra vanish in this block; more general quadratic Lindbladians
would add the corresponding $\theta$-dependent terms.  

\subsection{Realization-independent CP noise bound}

Collect the loss bath coupling vectors, including the passive ports, into
$C_\ell$ and the gain vectors into $C_g$.  Equation~\eqref{eq:s_langevin_io}
and $b_{\ell,{\rm out}}=b_{\ell,{\rm in}}+C_\ell^\dagger a$ give the
scattering matrices $S=I-C_\ell^\dagger\chi_\kappa C_\ell$ and
$T=C_\ell^\dagger\chi_\kappa C_g$ for annihilation and creation inputs.
The drift identity
$X_\kappa+X_\kappa^\dagger=C_gC_g^\dagger-C_\ell C_\ell^\dagger$
then yields commutator preservation,
\begin{equation}
 SS^\dagger-TT^\dagger=I .
 \label{eq:s_cp_commutator}
\end{equation}
For distinct ports $i,o$,
$S_{oi}=-\sqrt{\kappa_o\kappa_i}\,[\chi_\kappa]_{oi}$.
Independent vacuum gain inputs produce noise $(TT^\dagger)_{oo}$, so
\begin{equation}
 S_{{\rm out},o}^{\rm exc}(\omega)
 =(TT^\dagger)_{oo}
 =\sum_j|S_{oj}|^2-1
 \ge |S_{oi}|^2-1 .
\label{eq:s_cp_noise_bound}
\end{equation}
This standard amplifier bound \cite{Caves1982,CavesCombes2012,Clerk2010}
holds for any stable finite phase-preserving bosonic network with a
number-conserving quadratic Hamiltonian, linear gain/loss jumps, and
canonical Markov ports, including multiband and disordered chains.

\section{Pointwise local suppression and finite-rate constraints}
\label{sec:s_pointwise}
The loss matrix enters the excess above the positive-branch fixed-drift
floor through its projection on the measured response.  We first prove
that adjacent-cell jumps can eliminate this projection, then state the
rate conditions and the complementary sign branch.  This construction
addresses one selected frequency; a shared band is treated in the next
section.

\subsection{Adjacent-cell cancellation at one frequency}

Consider a nearest-neighbor chain with $d$ internal modes per cell.
At any real frequency let
\begin{equation}
 y=({\mathrm{i}}\omega I-X_\kappa)^{-\dagger}e_o,\qquad M=X+X^\dagger,\qquad
 \mathcal N_{\rm in}^{\rm exc}
 =\frac{y^\dagger\Gamma_g y}{\kappa_i|y_i|^2} .
 \label{eq:s_fixed_drift_noise}
\end{equation}
The unrestricted pointwise floor and sign-adaptive baths are known
\cite{LauClerk2018}.  The construction below enforces adjacent-cell support
for block-tridiagonal $M$.
Assume that $M$ is Hermitian and block tridiagonal, the resolvent exists,
$\kappa_i>0$, $y_i\ne0$, every cell block
$y_j\in\mathbb C^d$ is nonzero, and $y^\dagger M y>0$.  Among all CP
factorizations satisfying $\Gamma_g-\Gamma_\ell=M$, the minimum of
Eq.~\eqref{eq:s_fixed_drift_noise} is
\begin{equation}
\mathcal N_{\rm fixed}^{\min}
=\frac{y^\dagger M y}{\kappa_i|y_i|^2},
\label{eq:s_fixed_drift_min}
\end{equation}
and it is attained by a block-nearest-neighbor factorization.  Indeed,
\begin{equation}
y^\dagger\Gamma_g y
=y^\dagger M y+y^\dagger\Gamma_\ell y\ge y^\dagger M y .
\label{eq:s_fixed_drift_lower}
\end{equation}
Define the onsite projectors
$P_j=I_d-y_jy_j^\dagger/\|y_j\|^2$, write
$u_j=y_j/\|y_j\|$, and use geometrically balanced bond vectors
\begin{equation}
w_j=\sqrt{\frac{\|y_{j+1}\|}{\|y_j\|}}u_j
-\sqrt{\frac{\|y_j\|}{\|y_{j+1}\|}}u_{j+1},\qquad
K=\bigoplus_jP_j+\sum_{j=1}^{N-1}w_jw_j^\dagger ,
\label{eq:s_local_kernel}
\end{equation}
where the two entries of $w_j$ occupy cells $j$ and $j+1$.  Equivalently, in
the scalar case one may use
$v_j=(|y_jy_{j+1}|)^{-1/2}(y_{j+1}^*e_j-y_j^*e_{j+1})$.
The geometric-mean normalization depends only on neighboring response
ratios: multiplying $y$ by a common amplitude leaves $K$ unchanged.
Here $K\succeq0$ is block tridiagonal and
$\ker K=\operatorname{span}\{y\}$.  The Schur-complement argument in
Methods, Eq.~\textup{(22)}, applies to this block
kernel and gives a finite $\alpha$ with $M+\alpha K\succeq0$ on the
positive branch.  Since $Ky=0$, the local pair
$(\Gamma_g,\Gamma_\ell)=(M+\alpha K,\alpha K)$ attains
Eq.~\eqref{eq:s_fixed_drift_min}.  This existence statement does not
impose an arbitrary prescribed rate cap.

Both Gram matrices admit adjacent-cell jump factorizations even if singular.
For a PSD matrix with leading block $A$ and coupling $B$, positivity gives
$\operatorname{ran}B\subseteq\operatorname{ran}A$ and
$D-B^\dagger A^+B\succeq0$ on the remaining blocks.  Subtracting the PSD
matrix with blocks $A,B,B^\dagger,B^\dagger A^+B$ leaves that shorter
PSD remainder.  Block tridiagonality confines the subtracted factor to
the first two cells.  Iteration, with $A^+$ the Moore--Penrose inverse,
proves the required support.  Since $Ky=0$, the lower bound is saturated.

For arbitrary endpoint probe rates, the defining equation for $y$ gives
\begin{equation}
 \mathcal N_{\rm fixed}^{\min}-\mathcal N_{\rm Caves}
 =\frac{\kappa_o|y_o-\kappa_o^{-1}|^2}
 {\kappa_i|y_i|^2}
 =\frac{|\mathsf S_{oo}|^2}{|\mathsf S_{oi}|^2},
 \qquad
 \mathcal N_{\rm Caves}
 =1-\frac{1}{\kappa_i\kappa_o|y_i|^2},
 \label{eq:s_local_caves_relation}
\end{equation}
where $\mathsf S$ is the scattering matrix.  The finite-size excess above the
Caves floor is therefore precisely the output-reflection penalty.
At zero detuning in the amplification window, it decays as
$e^{-2\Lambda_{\rm resp}N}$.
The complementary sign branch, singular zero case, and finite-rate
penalty are derived in Sec.~\ref{sec:sign-adaptive-resource}.

\begin{figure}[t]
\centering
\includegraphics[width=0.90\textwidth]{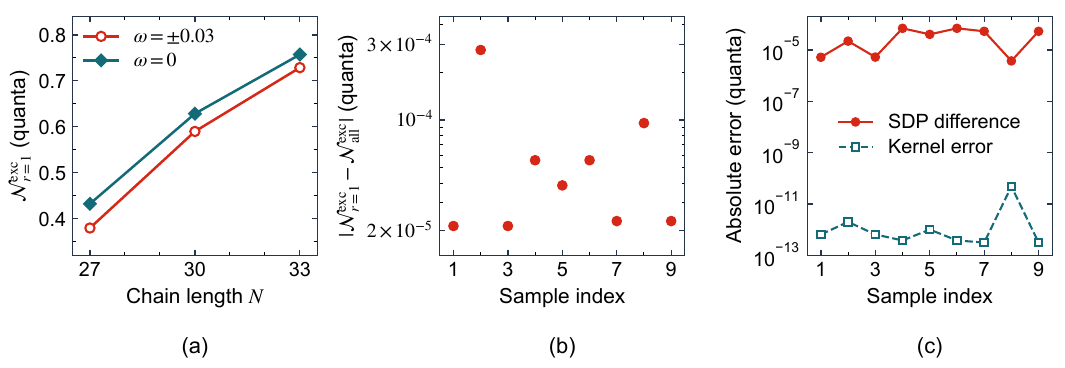}
\caption{Finite-frequency block-local optimality.
(a) Scalar nearest-neighbor optima at the three detunings defined in the
text.  (b) Absolute discrepancies between nearest-neighbor ($r=1$) and
unrestricted-range ($\mathrm{all}$) scalar SDPs, at most
$2.77\times10^{-4}$.  (c) Two-band SDP discrepancies, at most
$6.96\times10^{-5}$, and explicit-kernel errors relative to the analytic
optimum, below $8.44\times10^{-10}$.  Scalar explicit-kernel errors are
below $2.62\times10^{-12}$.  Sample indices order size before detuning;
these numerical discrepancies are distinct from a physical locality cost.}
\label{fig:s_finite_frequency_local_optimality}
\end{figure}

Zero-frequency semidefinite programs (SDPs) give largest $r=1$ versus
unrestricted discrepancies of $2.45\times10^{-7}$ at the high-gain
working points and $6.83\times10^{-6}$ across nine cases varying drift,
port rates and length.  The fitted reflection-penalty slope $-0.142$ agrees
with $-2\Lambda_{\rm resp}=-0.14157$.
Independent finite-frequency tests, also without a common cap, use
the same uniform scalar drift at each of $N=27,30,33$ and
$\omega=-0.03,0,0.03$.  The two-band samples have two modes per cell,
$8,9,10$ cells, and $\omega=-0.15,0,0.15$.  Their onsite, rightward, and
leftward drift blocks are, respectively,
\[
X_0=\begin{pmatrix}-1&-0.12{\mathrm{i}}\\-0.12{\mathrm{i}}&-0.9\end{pmatrix},\qquad
X_R=\operatorname{diag}(1.2,1.0),\qquad
X_L=\operatorname{diag}(0.10,0.08).
\]
Both tests use ports of rate $0.2$ on the first mode of the first and last
cells and satisfy $y^\dagger M y>0$.  Samples are indexed by increasing
size and then increasing frequency.  This two-orbital test is distinct
from the dimerized scalar chain in Sec.~\ref{sec:s_size_uniform}.
Supplementary Fig.~\ref{fig:s_finite_frequency_local_optimality} gives the SDP
discrepancies and explicit-kernel errors.  The scripts and sample
parameters are included in the data repository.
Its nine scalar size--detuning pairs differ from the nine
zero-frequency parameter cases above; their largest $r=1$ versus
unrestricted discrepancy is $2.77\times10^{-4}$.
For main-text Fig.~2, use the balanced
kernel defined above, with each marker tuned at $\omega_0=0$.
At $N=60$, numerical evaluation gives $\alpha=34.8609611$,
$\|\Gamma_g\|=134.623764$ and $\|\Gamma_\ell\|=139.520356$.
Hold this same reservoir pair fixed for the frequency scan over
$\omega\in[-0.05,0.05]$.
The exact port-damped response gives $\mathcal N(0)=0.994686305$
and $\mathcal N(0.05)=15.260255$, while the original bond reservoir gives
$197.657953$ and $209.651491$, respectively.  This numerical evaluation
checks positivity, drift and kernel residuals; it is distinct from the
rational complete-rate certificates used for the $R=8$ comparisons.
The previous max-normalized construction is retained in the archived
data but is not plotted in Fig.~2.

\subsection{Rate required for local cancellation}
Noise optimality and reservoir strength are distinct questions.
On the positive branch of the local construction,
define the minimum rate cap required for exact saturation by
\begin{equation}
 R_*(\omega)=
 \min_{\substack{\Gamma_\ell\succeq0,\ M+\Gamma_\ell\succeq0\\
                   \Gamma_\ell y=0,\ \Gamma_\ell\in\mathcal L_1}}
 \max\{\|\Gamma_\ell\|,\|M+\Gamma_\ell\|\},
 \label{eq:s_rate_star}
\end{equation}
where $\mathcal L_1$ denotes the Hermitian block-banded matrices satisfying
$(\Gamma)_{jk}=0$ whenever the cell distance $|j-k|>1$.  The
constraint $\Gamma_\ell y=0$ enforces exact saturation of the fixed-drift
noise bound.  We solve this convex problem by parameterizing every adjacent
cell Gramian in the null space of the corresponding response pair, avoiding
the ill conditioning of a generic SDP for exponentially localized $y$.
For the uniform chain at $N=64$, $\omega=0$, the saved solver estimate is
$R_*\simeq135.04$, with status \texttt{optimal\_inaccurate} and gain
minimum eigenvalue $-1.44\times10^{-7}$.  This is an approximate optimum,
not a certified feasible rate or lower bound.  The explicit balanced
construction supplies a separate feasible comparison.

Let $u_N=y_N/\|y_N\|$ and $P_N=I-u_Nu_N^\dagger$.  Suppose the balanced local
kernel and drift obey, with positive constants independent of $N$,
\begin{equation}
 K_Nu_N=0,\qquad
 k_0P_N\preceq K_N\preceq k_1I,\qquad
 \|M_N\|\le M_0,\qquad
 u_N^\dagger M_Nu_N\ge m_0>0 .
 \label{eq:s_uniform_rate_assumptions}
\end{equation}
Then the fixed-drift noise floor is attained by a block-local reservoir whose
gain and loss rates are bounded uniformly in $N$.  In particular, setting
\begin{equation}
 \alpha_0=\frac{M_0+M_0^2/m_0}{k_0},\qquad
 \Gamma_\ell=\alpha_0K_N,\qquad
 \Gamma_g=M_N+\alpha_0K_N,
 \label{eq:s_uniform_rate_construction}
\end{equation}
gives
\begin{equation}
 R_*(\omega)\le R_{\rm req}
 \equiv M_0+k_1\alpha_0 .
 \label{eq:s_uniform_rate_bound}
\end{equation}
To prove this, decompose $M_N+\alpha K_N$ with respect to
$\operatorname{span}\{u_N\}\oplus u_N^\perp$.  The lower-right block is at
least $(\alpha k_0-M_0)P_N$, while the off-diagonal block has norm at most
$M_0$.  For $\alpha=\alpha_0$, the Schur complement is therefore bounded below
by $m_0-M_0^2/(\alpha k_0-M_0)=0$.  Complete positivity and the rate bound then
follow from Eqs.~\eqref{eq:s_uniform_rate_construction} and
\eqref{eq:s_uniform_rate_assumptions}.

For the scalar chain the coercivity condition has a direct transfer-root
certificate.  Write $r_j=|y_{j+1}/y_j|$ and collect the balanced bond vectors
as the columns of $V_N$, so $K_N=V_NV_N^\dagger$.  The bond Gramian
$J_N=V_N^\dagger V_N$ has diagonal $(J_N)_{jj}=r_j+r_j^{-1}$ and
off-diagonal modulus $|(J_N)_{j,j+1}|=\sqrt{r_{j+1}/r_j}$.
Hence the nonzero spectrum of $K_N$ obeys the explicit Gershgorin bound
\begin{equation}
 \lambda_2(K_N)\ge\delta_N\equiv\min_j\left[
 r_j+r_j^{-1}-\sqrt{\frac{r_j}{r_{j-1}}}
 -\sqrt{\frac{r_{j+1}}{r_j}}\right],
 \label{eq:s_jacobi_margin}
\end{equation}
where absent endpoint terms are omitted.  At the uniform working point
and zero detuning $\omega=0$,
$r_1=1.3$ and the interior Riccati recursion is
$r_{j+1}=1.2-0.25/r_j$.  It decreases to the separated adjoint-profile root
$r_\infty=(1.2+\sqrt{1.2^2-1})/2=0.931662479\ldots$.
Substitution of the recursion in Eq.~\eqref{eq:s_jacobi_margin}, followed by
direct differentiation on $r_\infty\le r_j\le r_2$, gives
\begin{equation}
 \delta_N\ge r_\infty+r_\infty^{-1}-2=5.01256\times10^{-3} .
 \label{eq:s_scalar_uniform_gap}
\end{equation}
The endpoint rows have larger margins.  For $N\ge40$,
$G\ge G_{40}=3.8832$ and
$\|y\|^2/|y_1|^2\le1+r_1^2+(r_1r_2)^2/(1-r_3^2)=20.9716$.
The Caves relation then yields
$u_N^\dagger M_Nu_N\ge7.08\times10^{-3}$.  Together with
$\|M_N\|\le4.9$ and $\|K_N\|\le4.736$, Eq.~\eqref{eq:s_uniform_rate_bound} supplies a fully
explicit, although deliberately conservative, $N$-independent rate bound.
With the quoted constants this bound is $R_{\rm req}=3.21\times10^6$;
it proves finite uniform strength rather than a competitive device rate.
The explicit balanced construction instead approaches
$R_{\rm req}^{\rm bal}\simeq139.5$; this is distinct from the exact optimum
$R_*$ and far below the analytic upper bound in
Eq.~\eqref{eq:s_uniform_rate_bound}.

\subsection{Sign changes and the finite-rate noise penalty}
\label{sec:sign-adaptive-resource}

The local construction extends to both signs of the projected net gain.
Let $u=y/\|y\|$, $P=I-uu^\dagger$, $s=u^\dagger Mu$, and $b=PMu$.
The scalar $s$ is the net gain projected onto the measured response;
$b$ measures its coupling to orthogonal signal combinations.
For $\Gamma_g-\Gamma_\ell=M$ with both matrices PSD, positivity gives
$y^\dagger\Gamma_g y\ge\max\{y^\dagger My,0\}$ and hence
\begin{equation}
 \mathcal N_{\rm in}^{\rm exc}
 \geq
 \frac{\max\{y^\dagger M y,0\}}
 {\kappa_i|y_i|^2} .
\label{eq:s_sign_adaptive_floor}
\end{equation}

All matrix norms here are spectral (operator) norms, and vector norms are
Euclidean.  Take a block-nearest-neighbor kernel $K\succeq0$ with
$Ku=0$, $PKP\succeq k_0P$, and $k_0>0$.
For $s\neq0$, set $\sigma=\operatorname{sgn}(s)$.
On $\operatorname{span}\{u\}\oplus u^\perp$, the matrix
$A_\sigma=\sigma M+\alpha K$ has upper-left entry $|s|$, off-diagonal block $\sigma b$, and a lower block
bounded below by $(\alpha k_0-\|M\|)P$.  Hence the Schur complement is positive semidefinite
when
\begin{equation}
 \alpha\geq
 \frac{\|M\|+\|M\|^2/|s|}{k_0} .
\label{eq:s_sign_adaptive_alpha}
\end{equation}
Choose $(\Gamma_g,\Gamma_\ell)=(M+\alpha K,\alpha K)$ for $s>0$ and
$(\alpha K,-M+\alpha K)$ for $s<0$.  Both are local CP realizations,
preserve the drift, and attain Eq.~\eqref{eq:s_sign_adaptive_floor}.

At $s=0$, floor attainment requires both $\Gamma_g u=0$ and
$\Gamma_\ell u=0$, hence $Mu=0$.  If $Mu=0$, a sufficiently large local
$\alpha K$ attains zero noise.  If $Mu\ne0$, zero is only an infimum:
take $\Gamma_\ell=\alpha K+\epsilon I$ and
$\Gamma_g=M+\Gamma_\ell$, with $\epsilon>0$ and
$\alpha k_0\ge\|M\|+\|PMu\|^2/\epsilon$.
The Schur complement is PSD and $u^\dagger\Gamma_g u=\epsilon\to0$,
while the necessary rate diverges.  No finite-rate minimum is asserted at
this singular point.

There is a complementary rate lower bound.  Let
$R=\max\{\|\Gamma_g\|,\|\Gamma_\ell\|\}$.
If $s>0$, saturation of Eq.~\eqref{eq:s_sign_adaptive_floor} forces
$\Gamma_\ell u=0$, so the gain matrix has the positive block form
\begin{equation}
 \Gamma_g=\begin{pmatrix}s&b^\dagger\\ b&C\end{pmatrix}\succeq0 .
\end{equation}
The $2\times2$ positivity condition implies
$\|b\|^2\leq s\|C\|\leq sR$.  For $s<0$ the same argument applies to
$\Gamma_\ell$ and gives
\begin{equation}
 R\geq \frac{\|PMu\|^2}{|s|} .
\label{eq:s_rate_resource_lower_bound}
\end{equation}
Therefore, if $s(\omega_*)=0$ while $PMu(\omega_*)\neq0$, no finite-rate CP
reservoir can attain the pointwise optimum at $\omega_*$.  The required rate
diverges at least as $|s|^{-1}$ on approaching the crossing.  This lower bound
does not use locality; Eq.~\eqref{eq:s_sign_adaptive_alpha} supplies a matching
local construction up to a model-dependent factor.
Finite transverse coupling $b$ therefore requires increasing reservoir
strength as the projected net gain vanishes.

For any finite cap $R>0$ with
$R\geq\max\{\|\Gamma_g\|,\|\Gamma_\ell\|\}$, define
$\delta_R=u^\dagger\Gamma_g u-\max\{s,0\}$.
Every compatible CP realization satisfies
\begin{equation}
 \delta_R\geq
 \frac{\bigl[\|b\|^2-R|s|\bigr]_+^2}{4R\|b\|^2},
 \qquad b=PMu.
\label{eq:s_finite_rate_penalty_unit}
\end{equation}
Restoring the input normalization gives
\begin{equation}
 \mathcal N_{\rm in}^{\rm exc}
 \geq
 \frac{\|y\|^2\max\{s,0\}}{\kappa_i|y_i|^2}
 \;+\;
 \frac{\|y\|^2}{\kappa_i|y_i|^2}
 \frac{\bigl[\|b\|^2-R|s|\bigr]_+^2}{4R\|b\|^2}.
\label{eq:s_finite_rate_penalty}
\end{equation}
The penalty term is set to zero when $b=0$ in both
Eqs.~\eqref{eq:s_finite_rate_penalty_unit} and \eqref{eq:s_finite_rate_penalty}.
For $s>0$, write $a=u^\dagger\Gamma_\ell u$, so that
$u^\dagger\Gamma_g u=s+a$.  Positivity and the cap imply
$\|P\Gamma_g u\|\leq\sqrt{R(s+a)}$ and
$\|P\Gamma_\ell u\|\leq\sqrt{Ra}$.
Since $b=P(\Gamma_g-\Gamma_\ell)u$, the triangle inequality gives
$\|b\|\leq\sqrt{R(s+a)}+\sqrt{Ra}$.
Solving this inequality for $a$ gives
$a\geq[\|b\|^2-Rs]_+^2/(4R\|b\|^2)$.  For $s<0$, set
$a=u^\dagger\Gamma_g u$ and exchange gain and loss; the same calculation
replaces $s$ by $|s|$.  Thus at a sign crossing,
$\delta_R\geq\|b\|^2/(4R)$: a finite reservoir rate produces a finite,
quantitative excess above the pointwise floor.  This statement is independent
of locality and therefore also applies to the block-local constructions.

\section{Shared-band optimization with a complete matrix rate cap}
\label{sec:s_rate_capped_band}

A shared device keeps every reservoir coupling fixed during a frequency
scan.  Different frequencies probe different spatial combinations of
reservoir noise, and the objective is their band average.
The complete-rate comparison separates pointwise redesign from one
shared reservoir pair and yields the finite-chain performance plotted
in Fig.~3.

We first consider the uniform chain at $N=40,48,64$, with
$t_R=1$, $t_L=1/4$, $\gamma=6/5$, fixed ports
$\kappa_i=\kappa_o=1/5$, and $\mathcal B=[-1/20,1/20]$.
Each optimization uses one frequency-independent
pair of reservoir matrices.  The tested complete operator-norm budgets are
$R=8,32,256$, and the allowed matrix ranges are $r=1,2$ and unrestricted.

\subsection{Exact polynomial integration and the gain minimum}
Write $h_j=y_j/y_1$, indexing sites from one.  All but the final resolvent
equation have zero source, giving
\begin{equation}
h_1=1,\quad h_2=13/10-{\mathrm{i}}\omega,\quad
h_{j+1}=(6/5-{\mathrm{i}}\omega)h_j-h_{j-1}/4.
\label{eq:s_polynomial_profile}
\end{equation}
The right endpoint fixes the common amplitude $y_1$ and does not alter
these polynomials.  Methods, Eq.~\textup{(25)},
gives their exact rational Legendre integral.  For independent
Gauss--Legendre quadrature at orders $N+8$ and $2N+9$, the maximum
absolute entrywise difference from the rational Gram matrix is below
$1.49\times10^{-14}$ in the tested sizes.  Orders at least $N$ integrate
the polynomial entries exactly in real arithmetic; the discrepancy is
numerical roundoff rather than quadrature truncation error.

The negative probed drift $-X_\kappa$ is diagonally similar to a real
symmetric tridiagonal matrix $A$, whose diagonal entries are $6/5$ with
endpoint increments $1/10$ and whose off-diagonal entries have magnitude
$1/2$.  Its eigenvalues $a_j$ are positive; the resolvent is correspondingly
similar to $({\mathrm{i}}\omega I+A)^{-1}$.  Hence
$G_N(\omega)=(1/25)/\prod_j(a_j^2+\omega^2)$.
The product increases with $|\omega|$.  The whole-band minimum is therefore
at $\pm\Omega$, with no frequency-grid assumption.  The checker evaluates
that endpoint determinant with exact rational real and imaginary parts.
The resulting minima at $N=40,48,64$ are $2.65508494$, $7.60125245$ and
$62.30151821$; the entire band lies on the positive fixed-drift branch.

\subsection{Unrestricted shared-band benchmark without a rate cap}
For $W\succ0$, set $A=W^{1/2}MW^{1/2}$ and
$F=W^{1/2}\Gamma_gW^{1/2}$.  Without locality or a cap,
$\min\operatorname{Tr}F$ subject to $F\succeq0$, $F-A\succeq0$
has value $\operatorname{Tr}A_+$: projection onto the positive eigenspace
gives the lower bound, and $F=A_+$ attains it.  Thus
$\mathcal J_{\rm sh}^{\infty}=\operatorname{Tr}A_+$, realized by
$\Gamma_g=W^{-1/2}A_+W^{-1/2}$.
This standard congruence construction may be ill conditioned or exceed
a finite cap; it only supplies the uncapped value benchmark.

On a band with $y^\dagger My>0$,
$\overline f=\operatorname{Tr}(WM)=\operatorname{Tr}A$, so the sharing
cost is $\operatorname{Tr}A_-$, where $A_-=(-A)_+$.
Congruence preserves inertia, making this cost positive whenever $M$
has a negative eigenvalue.  Both chains satisfy this condition because
$\operatorname{Tr}M=-2\sum_j\gamma_j<0$, where $\gamma_j>0$ is the
on-site amplitude damping.  In physical terms, zero excess would require
$\operatorname{Tr}(W\Gamma_\ell)=0$.  Since $W\succ0$, that forces
$\Gamma_\ell=0$ and hence $\Gamma_g=M$, contradicting gain positivity.
This argument gives finite-length strictness; the next section derives
quantitative bounds independent of length.

For a feasible capped problem on this positive branch, define
\begin{equation}
\begin{aligned}
C_{\rm sh}&=\mathcal J_{\rm sh}^{\infty}-\overline f
             =\operatorname{Tr}A_-,\\
C_{\rm cap}&=\mathcal J^*_{N,R,\mathrm{all}}
              -\mathcal J_{\rm sh}^{\infty},\\
C_{\rm loc}^{(r)}&=\mathcal J^*_{N,R,r}
                    -\mathcal J^*_{N,R,\mathrm{all}}.
\end{aligned}
\label{eq:s_shared_cost_decomposition}
\end{equation}
The excess obeys
$\mathcal J^*_{N,R,r}-\overline f=C_{\rm sh}+C_{\rm cap}+C_{\rm loc}^{(r)}$.
Each term is nonnegative by inclusion of the feasible sets.
The first measures frequency sharing without a rate cap, the second
the additional restriction of a finite cap, and the third the restriction
of range at that same cap.  A finite-cap optimum minus the adaptive floor
therefore measures the combined sharing and cap cost; it does not
determine the two contributions separately.

\subsection{A certified frequency-adaptive benchmark under the same cap}
At each frequency independently, set $u=h/\|h\|$ and
\begin{equation}
\Gamma_\ell=7(I-uu^\dagger),\qquad \Gamma_g=M+\Gamma_\ell.
\label{eq:s_nonlocal_pointwise_family}
\end{equation}
Since $M+7I\succ0$, gain positivity is equivalent to
$q_N(x):=h^\dagger[I-7(M+7I)^{-1}]h\geq0$.
The polynomial coefficients are rational.  On $|x|\le1$, the bound
$q_N(0)-\sum_{k\ge1}|[x^k]q_N(x)|$ is at least
$0.00300010597$, $0.00579268675$, and $0.00710883906$
at $N=40,48,64$, respectively.
The loss norm is 7 and $\lambda_{\max}(M)<0.1$, so the positive gain
matrix has norm below 7.1.  This family annihilates $y$ on the loss side
and attains the pointwise floor at every frequency at all tested caps.
Its matrices change with frequency; it is a benchmark of separately
designed devices and not a shared physical reservoir.

\subsection{Primal and dual finite-matrix certificates}
Methods, Eqs.~\textup{(24)}--\textup{(27)},
give the normalized optimization, feasibility repair and
residual-corrected dual lower bound.  In the saved certificates,
binary-float entries and factors are interpreted as exact dyadic
rationals.  Write $L=\Gamma_\ell$; the gain is defined by $M+L$
with rational $M$.
All tested budgets exceed $\|M\|$, making the stated interior anchor
feasible.  The smallest verified PSD margin across the 27 cases is
$9.65\times10^{-10}$.  For $r=1,2$, the verified banded Cholesky factors
give jumps on at most $r+1$ consecutive sites.

After projection onto the allowed matrix entries, a candidate may
violate positivity by at most $v_{\rm viol}$ in its worst constraint.
The normalized anchor $\widetilde L_0=(I-\widetilde M)/2$ has margin
$\mu_0=(1-\|M\|/R)/2>0$ in all four constraints.
Mixing an anchor fraction $\theta\ge v_{\rm viol}/(\mu_0+v_{\rm viol})$
therefore repairs the candidate.  Both the projection and anchor
preserve the allowed range.  For each repaired constraint $T$, a factor
of $T-\delta_T I$ is recovered numerically.  Its stored entries are
then treated as exact dyadic rationals: the Gram product is positive
semidefinite exactly, and the row-sum residual in main-text
Eq.~\textup{(26)} gives a strict positive margin.
Solver feasibility tolerances are not used as proof of positivity.

The dual lower bound follows from an exact Lagrange identity.
With the four positive factors $A_d,B_d,C_d,D_d$ and allowed-entry
residual $E_d$ defined in Methods,
\begin{equation}
\begin{aligned}
\operatorname{Tr}(W\widetilde L)={}&
\operatorname{Tr}[(D_d-B_d)\widetilde M]-\operatorname{Tr}(C_d+D_d)
+\operatorname{Tr}(E_d\widetilde L)\\
&+\operatorname{Tr}(A_d\widetilde L)
+\operatorname{Tr}[B_d(\widetilde L+\widetilde M)]\\
&+\operatorname{Tr}[C_d(I-\widetilde L)]
+\operatorname{Tr}[D_d(I-\widetilde L-\widetilde M)].
\end{aligned}
\label{eq:s_dual_identity_detail}
\end{equation}
The last four traces are nonnegative.  Range projection does not
change the residual trace because forbidden entries of $\widetilde L$
vanish.  Since $0\preceq\widetilde L\preceq I$, the remaining trace is
bounded below by $-N\|E_d\|_\infty$.  Exact rational products and
outward rounding give the bound in main-text
Eq.~\textup{(27)}, in the same noise units as the
feasible upper bound after restoring $R$ and $\operatorname{Tr}(WM)$.

Supplementary Table~\ref{tab:s_finite_band_comparison} uses the normally
ordered input-referred noise convention.

\begin{table}[htbp]
\centering
\begin{tabular}{c c c c c}
\hline
$N$ & $R$ & Nonlocal noise & $r=1$ noise & Certified $r=1$ minus nonlocal \\
\hline
40 & 8 & $2.293032$ & $3.357281$ & $[1.0642486,1.0642504]$ \\
48 & 8 & $2.349272$ & $3.413642$ & $[1.0643695,1.0643709]$ \\
64 & 8 & $2.374013$ & $3.438565$ & $[1.0645489,1.0645542]$ \\
40 & 32 & $2.293031$ & $2.326321$ & $[0.0332879,0.0332926]$ \\
48 & 32 & $2.349274$ & $2.391660$ & $[0.0423815,0.0423903]$ \\
64 & 32 & $2.374012$ & $2.420524$ & $[0.0465081,0.0465146]$ \\
64 & 256 & $2.374003$ & $2.417108$ & $[0.0430805,0.0431291]$ \\
\hline
\end{tabular}
\caption{Certified shared-band comparisons at finite chain length.
Noise entries are rounded interval centers; the last column encloses
the nearest-neighbor minimum minus the unrestricted-range minimum.
The chain length $N$ and complete matrix rate cap $R$ are listed for each case.}
\label{tab:s_finite_band_comparison}
\end{table}
Noise columns are rounded interval centers; only the interval columns
are bound statements.  The full CSV reports all 27 problems, including
$r=2$.  The largest reported optimum interval width is
$3.0362\times10^{-5}$, well below the resolved locality differences.
At $N=64,R=8$, the pointwise floor average is $0.99622518$ and the
combined cost of frequency sharing and the finite rate cap relative to
this floor is between $1.37778758$ and $1.37778873$.
The original bond reservoir obeys the same cap and has band-averaged
noise $201.59135$.

\section{Length-independent noise bounds and design targets}
\label{sec:s_size_uniform}
\label{sec:s_targets}

The first $m$ sites, called the input-side prefix, have
length-independent normalized noise weights.
Its constrained minimum gives a lower bound after a controlled crossing
correction; a reservoir with a reserved bridge rate gives an achievable
upper bound at every longer finite length.  Their strict separation
establishes the locality cost.  In the proofs, $G=\Gamma_g$ and
$L=\Gamma_\ell$ denote the gain and loss rate matrices; $G_N(\omega)$
denotes the transmission power gain.

\subsection{Length-independent boundary response}
Use zero-based site indices in this section: sites $0$ and $N-1$
correspond to main-text sites $1$ and $N$.  Methods,
Eq.~\textup{(28)}, gives the boundary recurrence
for period-one and period-two Jacobi drifts with positive directed
couplings.  Relabeling that recurrence gives $h_j=y_j/y_0$ here.
The first $N-1$ adjoint resolvent equations have zero source; only the
last determines $y_0$.  Thus the normalized response is exactly
independent of $N$ for $N>j$ and
\begin{equation}
W_{ij}(\Omega)=\frac1{2\Omega\kappa_i}\int_{-\Omega}^{\Omega}
h_i(\omega)h_j(\omega)^*\,{\mathrm{d}}\omega
\end{equation}
defines consistent principal blocks at every length.

Conjugation averaging gives real symmetric reservoirs on this symmetric
band without changing drift, noise, PSD, rate, or matrix range.

The two certified drifts have $\kappa_i=\kappa_o=1/5$ and
\begin{equation}
\begin{array}{c|ccc}
 & (\gamma_0,\gamma_1)&(t_{R,0},t_{R,1})&(t_{L,0},t_{L,1})\\ \hline
\text{uniform}&(6/5,6/5)&(1,1)&(1/4,1/4)\\
\text{dimerized}&(6/5,23/20)&(23/20,17/20)&(3/20,7/20).
\end{array}
\label{eq:s_two_jacobi_models}
\end{equation}
The second is a two-sublattice chain with alternating onsite loss
and hopping.  Range $r$ counts physical sites, not two-site cells.

\subsection{Noise bound from a fixed input-side region}
For the compatible Gram and drift blocks, the prefix lifting inequality
is Methods, Eq.~\textup{(29)}.  The constants in
Eq.~\textup{(30)} give the affine noise bound in
main-text Eq.~\textup{(13)}.  With zero-based
indices the cross-boundary sum is
$b_{m,r}=\sum_{i<m\le j,\ j-i\le r}|W_{ij}|$.
Including absent cross pairs when $N<m+r$ only enlarges the correction.
The following argument supplies the full positivity identity behind
the bound, with the same constants after the index shift.

For $R=0$, feasibility forces $G=L=M_N=0$ and the result is immediate.
For $R>0$ and $0\preceq G\preceq RI$, the Hermitian contraction $2G/R-I$ has
every off-diagonal entry of modulus at most one.  Hence $|G_{ij}|\le R/2$.
The prefix principal blocks satisfy the same finite constraints, and the
tail contribution $\operatorname{Tr}(W_{\rm tail}G_{\rm tail})$ is
nonnegative since both factors are PSD.  Each allowed cross pair adds
$2\operatorname{Re}(W_{ij}G_{ji})\ge-R|W_{ij}|$.  This proves
Eq.~\textup{(29)}, including the bond at the witness endpoint.

For the prefix, write $G=L+M_m$.  The exact identity is
\begin{equation}
\begin{aligned}
\operatorname{Tr}(W_mG)={}&\alpha_m-R\operatorname{Tr}(C+D)
+\operatorname{Tr}(EL)\\
&+\operatorname{Tr}(AL)+\operatorname{Tr}(BG)\\
&+\operatorname{Tr}[C(RI-L)]+\operatorname{Tr}[D(RI-G)].
\end{aligned}
\end{equation}
The last four terms are nonnegative by positivity of the corresponding
constraint and dual matrices.
The range projection is legitimate because forbidden entries of $L$
vanish.  Using $0\le L_{ii}\le R$ and $|L_{ij}|\le R/2$ yields
$\operatorname{Tr}(EL)\ge-R\Lambda(E)$.  Combine the two bounds and
nonnegative noise to obtain Eq.~\textup{(13)}.
The dual factors need not be stationary or optimal at the new budget.

At rational bandwidth, all entries of $W_m$ are rational Legendre sums.
Stored binary-float factors are interpreted as exact dyadic rationals,
so their Gram products are PSD exactly and the displayed residual penalty
is evaluated exactly.  The reported lower constants are rounded downward
and slopes upward.  For fixed factors, $\alpha_m(\Omega)$ is an even
polynomial; $\beta_{m,r}(\Omega)$ consists of finite polynomial absolute
values and positive parts.  They are explicit functions of bandwidth and
range, with no $N$ dependence.  For $\beta_{m,r}>0$, target noise $T$
requires $R\ge[\alpha_m-T]_+/\beta_{m,r}$.  This necessary condition
does not establish saturation or a universal optimal scaling law.

\subsection{Extending the reservoir within the rate budget}
Methods specifies the prefix, tail bond pairs and onsite loss used in
the extension.  The complete-budget condition is
Eq.~\textup{(31)}; in the zero-based notation here,
$b_j=t_{R,j}+t_{L,j}$ for every bond, and the bridge is bond $m-1$.
The bridge is omitted at $N=m$.
We verify below that this prescription preserves the drift, positivity
and complete rates.

Every added matrix is a positive rank-one term.  The gain/loss bond
diagonals cancel in their difference, their off-diagonal difference is
$b_j$, and onsite losses supply $-2\gamma_j$.  This includes the bridge.
Thus $G-L=M_N$, so the prescribed drift is preserved.
Absolute row sums bound the tail gain norm
by $b_0+b_1$ and tail loss norm by $2\gamma_{\max}+b_0+b_1$.
Each bridge matrix, gain and loss, has operator norm $b_{m-1}$.
The norm of a direct sum is the larger block norm, so the triangle
inequality bounds each complete matrix by $R_0$ after the bridge is
added.  All added terms are nearest-neighbor or onsite.

For $m=64,R_0=8$, the uniform bridge rate is $5/4$ and the dimerized one
is $6/5$.  Prefix caps are therefore $27/4$ and $34/5$.
Both tails have gain/loss caps 2.5/4.9, so
Eq.~\textup{(31)} holds.  Arbitrary prefix matrices
require range at most 63, independent of device length.  Range-two
prefixes retain range two after extension, and their banded Cholesky
factors give jumps on at most three consecutive sites.

\subsection{Bounding the noise from the added sites}
For the recurrence in Eq.~\textup{(28)}, the two-site
Floquet roots satisfy $\lambda^2-\tau(\omega)\lambda+d=0$, with
\begin{equation}
\begin{aligned}
d&=\frac{t_{L,0}t_{L,1}}{t_{R,0}t_{R,1}}>0,\\
\tau(\omega)&=\frac{(\gamma_0-{\mathrm{i}}\omega)(\gamma_1-{\mathrm{i}}\omega)}{t_{R,0}t_{R,1}}
-\frac{t_{L,0}}{t_{R,1}}-\frac{t_{L,1}}{t_{R,0}}.
\end{aligned}
\end{equation}
For $q>\sqrt d$, the circle $|\lambda|=q$ maps under
$\lambda+d/\lambda$ to an ellipse.  A trace inside that ellipse has both
roots inside the circle: the interior is connected, at trace zero both
roots have modulus $\sqrt d$, and a root can exit only at the ellipse.
Choose a rational $q<1$ and rational $T,B,a>0$ such that on the band
$|\operatorname{Re}\tau|\le T$, $|\operatorname{Im}\tau|\le B$,
$\operatorname{Re}\tau\ge a$, and
\begin{equation}
\frac{T^2}{(q+d/q)^2}+\frac{B^2}{(q-d/q)^2}<1,\qquad
a^2-B^2-4d\ge\sigma^2>0.
\label{eq:s_floquet_rational_conditions}
\end{equation}
The principal square root of $\tau^2-4d$ has real part at least
$\sigma$, since its squared real part is
$(|\tau^2-4d|+\operatorname{Re}(\tau^2-4d))/2$.
Consequently the roots are separated by at least $\sigma$, and the root
of smaller modulus has modulus at most
\begin{equation}
s=\frac{2d}{a+\sigma}<1.
\end{equation}
For each parity $p=0,1$, let $H_p\ge|h_p|$ and $H_{p+2}\ge|h_{p+2}|$
be rational triangle bounds on the first recurrence values.  The root
decomposition of $h_{2k+p}$ gives
\begin{equation}
|h_{2k+p}|\le P_pq^k+Q_ps^k,\qquad
P_p=\frac{H_{p+2}+sH_p}{\sigma},\quad
Q_p=\frac{H_{p+2}+qH_p}{\sigma}.
\label{eq:s_floquet_profile_envelope}
\end{equation}
Summing this envelope gives the rational geometric-tail bound in
Methods, Eq.~\textup{(33)}, after the one-site index
shift.  The explicit noise ceiling in
Eq.~\textup{(34)} uses Legendre Gram entries through
cutoff $c=160$ and the remaining tail norm.
It holds for every finite $N\ge m$: chains ending before the cutoff
omit positive bond terms, while longer chains are bounded by the
geometric sum.  No infinite-chain stable steady state is assumed.

\subsection{Stability and gain throughout the band}
The sufficient stability margin
\begin{equation}
\gamma_{\min}-\sqrt{t_{R,0}t_{L,0}}-\sqrt{t_{R,1}t_{L,1}}>0
\end{equation}
makes the probed drift diagonally similar to a negative-definite symmetric
Jacobi matrix.  If its positive counterpart has eigenvalues $a_n>0$,
\begin{equation}
G_N(\omega)=\frac{\kappa_i\kappa_o(\prod_{j=0}^{N-2}t_{R,j})^2}
{\prod_{n=1}^N(a_n^2+\omega^2)}.
\label{eq:s_jacobi_gain}
\end{equation}
The minimum on the entire band is at $\pm\Omega$.  Its endpoint can be
evaluated rationally via the last normalized resolvent equation
\begin{equation}
D_N=(\gamma_{N-1}+\kappa_o/2-{\mathrm{i}}\Omega)h_{N-1}
-t_{L,N-2}h_{N-2},\qquad G_N(\Omega)=\frac{\kappa_i\kappa_o}{|D_N|^2}.
\end{equation}
The envelope in Eq.~\eqref{eq:s_floquet_profile_envelope} bounds $D_N$
for all $N$ beyond an even cutoff.  The remaining lengths from 64 to
that cutoff are checked individually with exact rational arithmetic.
The all-length whole-band gain lower bounds, at halfwidths 0.03, 0.05, 0.07,
are respectively 92.54667, 62.30151, 34.97199 for the uniform drift and
101.86724, 69.78286, 40.10903 for the dimerized drift.  Exact finite checks
end before lengths 86, 88, 94, respectively; the envelope covers all
remaining lengths.  Odd lengths are included.

\subsection{Rate, bandwidth, and minimum-range targets}

The following displayed lower bounds are rounded down and upper bounds
up, independently for each column.  Gap bounds are computed from the
unrounded certificate constants before rounding.  All rows hold for
every finite $N\ge64$ and $R=8$; $U$ is a
construction upper bound, not an optimized infinite-length value.
\begin{table}[htbp]
\centering
\begin{tabular}{c c c c c c}
\hline
Model & $\Omega$ & Noise upper $U$ & $r=1$ noise lower & $r=1$ gap lower & $r=2$ gap lower\\
\hline
uniform & 0.03 & 1.797653 & 3.283474 & 1.485822 & 0.172717\\
uniform & 0.05 & 2.410804 & 3.423463 & 1.012659 & 0.007523\\
uniform & 0.07 & 2.948057 & 3.637578 & 0.689522 & unresolved\\
dimerized & 0.03 & 2.198722 & 2.993226 & 0.794505 & 0.093375\\
dimerized & 0.05 & 2.535919 & 3.137701 & 0.601782 & unresolved\\
dimerized & 0.07 & 2.913504 & 3.363965 & 0.450462 & unresolved\\
\hline
\end{tabular}
\caption{Certified noise and locality bounds for every finite $N\ge64$
at complete matrix cap $R=8$.  The achievable noise ceiling $U$ is
compared with the nearest-neighbor lower bound.  The last two columns
bound the excess over the unrestricted-range optimum for ranges one
and two.  Each row uses its listed band halfwidth $\Omega$.}
\label{tab:s_all_length_noise}
\end{table}
The locality gap lower bound is the local lower bound minus $U$, hence
also bounds $\mathcal J^*_{N,8,r}-\mathcal J^*_{N,8,\mathrm{all}}$.
Negative witnesses in the three unresolved range-two cases do not imply
a vanishing or negative true gap.

At $\Omega=0.05$, the uniform nearest-neighbor constants are
$\alpha>4.88099438160254$, $\beta<0.182191357024353$.
Thus achieving the fixed certified ceiling $U+0.1$ requires
$R>13.00934766944$.  The dimerized constants give $R>10.58946231157$
for the analogous target.  These are necessary rates for a fixed
achievable noise ceiling; they are not sufficient rates or bounds on
total pump power.

Two additional strictly certified range-two prefixes at the reserved
caps extend by the construction in Eq.~\textup{(31)}.  Their noise ceilings are
2.573275 (uniform) and 2.862384 (dimerized).
It follows that at $R=8$, $\Omega=0.05$, for every $N\ge64$, matrix
range two is necessary and sufficient for a noise target 2.6 in the
uniform model and 2.9 in the dimerized model.  Necessity follows from
the nearest-neighbor lower bounds; sufficiency is the explicit
range-two extension.  This statement concerns those prescribed ceilings,
not distance from the unknown nonlocal optimum at each length.

\subsection{Continuous bandwidth certificates for fixed reservoirs}
Methods, Eq.~\textup{(35)}, gives continuous
bandwidth bounds from positivity of the fixed-device noise integrand.
For each model, keep the reference and range-two prefixes designed at
$\Omega=0.05$ unchanged, including the reserved bridge budget.
The witness set contains 81 halfwidths from 0.03 to 0.07 in steps of
0.0005, and $U^q_m(b)$ uses these same fixed prefixes.

Here $U^q_m(b)$ is the explicit-extension bound in Eq.~\textup{(34)} for
the same fixed prefix at every width.  Partition $[0.03,0.07]$ into
160 cells of width 0.00025 and evaluate all terms rationally at $R=8$.
The minimum certified differences $L_{[a,b]}(8)-U^{\rm ref}_{[a,b]}$
exceed 0.47876 (uniform) and 0.29893 (dimerized).
The whole-band gain bounds at 0.07 cover every smaller band.
For targets $T=2.6$ and $T=2.9$, respectively,
$U^2_{[a,b]}<T<L_{[a,b]}(8)$ holds on every cell covering
$[0.03,0.0505]$ and $[0.03,0.0515]$.
Thus the minimum integer matrix range is exactly two on these intervals,
at every finite $N\ge64$.  Neither prefix is retuned with bandwidth.
The interval endpoints are sufficient certified coverage; larger widths
are not excluded by failure of this construction bound.

\subsection{Length-independent noise-error bounds at the nominal drifts}
\label{sec:s_all_length_tolerance}
For the two nominal Markov chains at $\Omega=0.05$, the response
vector in Eq.~\textup{(41)} is the one defining
$W_N$.  Hence its band-averaged energy is $\operatorname{Tr}W_N$.
Compatibility fixes the first 158 weights at every longer length,
and the rational Floquet envelope bounds the remaining sum.
For shorter $N\ge64$, positivity bounds the energy by the same truncated
sum plus tail.  This gives $\operatorname{Tr}W_N\le C_Q$ independent
of $N$, without assuming an infinite-chain steady state.
Extend new range-two prefixes of cap $6.5$ using Eq.~\textup{(31)};
their complete rates are bounded by $7.75$ (uniform) and $7.70$ (dimerized).
Subtract their certified noise ceilings from the existing
nearest-neighbor lower bounds at cap $8$ to obtain $\Delta_0$.
The exact polynomial integrals and tail bounds give the conservative
values below.
\begin{table}[htbp]
\centering
\begin{tabular}{l c c c c}
\hline
Model & Local noise lower & Reference upper & $C_Q$ upper &
$\Delta_{\rm res}$ lower\\
\hline
uniform & $3.423463$ & $2.598690$ & $83.368629$ & $0.683090$\\
dimerized & $3.137701$ & $2.962589$ & $75.249704$ & $0.037487$\\
\hline
\end{tabular}
\caption{Length-independent noise-error budgets for the two nominal
Markov chains.  The remaining resolvable gap is bounded at combined force-noise
error $\delta=0.0005$ and measurement error $u_{\rm meas}=0.05$ per device.}
\label{tab:s_nominal_noise_tolerance}
\end{table}
\FloatBarrier
The last column uses combined force-noise error $\delta=0.0005$
and per-device measurement error $u_{\rm meas}=0.05$ in
Eq.~\textup{(41)}, evaluated before rounding.
It remains positive for every finite $N\ge64$, conditional on retained
response matching and physical feasibility.
This result concerns force-noise errors at nominal Markov drifts;
it does not extend the continuous finite-auxiliary optimality grid to
all lengths or certify arbitrary drift or pump errors.

\section{Microwave realization and noise tolerances}
\label{sec:s_microwave}

The proposed apparatus in main-text Fig.~1(a)
uses Josephson mixers for phase-controlled conversion and pair creation
\cite{Sliwa2015,Lecocq2017,RanzaniAumentado2015}, with calibrated
cryogenic scattering and noise detection \cite{LecocqConversion2016,Bernier2017}.
For component scales, Ref.~\cite{Lecocq2017} reports modes at
$(4.155,5.756,7.915)$ GHz, energy linewidths $(29.8,29.5,59.3)$ MHz,
and temperature $20$ mK.  These few-mode experiments support the design
scales, but the proposed array has not been demonstrated.

Let $\nu_j,\nu_b$ denote the laboratory angular frequencies of signal
and auxiliary modes.  Pumps at $|\nu_b-\nu_j|$ and $\nu_b+\nu_j$
select conversion and pair creation.  On each baseline bond $(j,j+1)$,
independent auxiliary annihilation operators $b_{\ell,j},b_{g,j}$ label
loss and gain.  Each auxiliary couples to either signal with complex
pump-controlled amplitude $g$ and loses energy to a near-vacuum
termination at rate $\kappa_b$, corresponding to amplitude decay
$\kappa_b/2$.  The rotating-frame interactions are
\begin{equation}
\begin{aligned}
\widehat H_{\ell,j}&=g b_{\ell,j}^\dagger(a_j-a_{j+1})+\mathrm{h.c.},\\
\widehat H_{g,j}&=g b_{g,j}^\dagger(a_j^\dagger+a_{j+1}^\dagger)
+\mathrm{h.c.}
\end{aligned}
\label{eq:s_auxiliary_bond}
\end{equation}
For either auxiliary $b$, define $F=a_j-a_{j+1}$ for loss and
$F=a_j^\dagger+a_{j+1}^\dagger$ for gain.  The interaction is then
$g b^\dagger F+\mathrm{h.c.}$.  With $b_{\rm in}$ the incident vacuum field,
the auxiliary Langevin equation and its leading adiabatic solution are
\begin{equation}
\begin{aligned}
\dot b&=-\frac{\kappa_b}{2}b-{\mathrm{i}} gF+\sqrt{\kappa_b}\,b_{\rm in},\\
b_{\rm sig}&\simeq-\frac{2{\mathrm{i}} g}{\kappa_b}F,
\qquad L_{\rm eff}=\frac{2|g|}{\sqrt{\kappa_b}}F.
\end{aligned}
\label{eq:s_auxiliary_elimination}
\end{equation}
Here $b_{\rm sig}$ is the signal-induced part of the auxiliary field;
the vacuum drive remains in the effective reservoir.  The effective jump
$L_{\rm eff}$ is specified up to a global phase.
The probe detuning $\omega$ is measured from the local signal carrier
in the rotating frame.  For resonant pumps, $|g|,|\omega|\ll\kappa_b$,
and auxiliary relaxation faster than the signal dynamics, elimination gives
Eq.~\eqref{eq:s_bond_jumps}, with $c=4|g|^2/\kappa_b$ to leading
Markov order \cite{GardinerZoller2004,MetelmannClerk2015}.
The factor four comes from the auxiliary amplitude decay rate
$\kappa_b/2$ and the output coupling $\sqrt{\kappa_b}$.
The loss minus sign is a relative phase $\pi$, not a
negative damping rate.  The figure labels effective coupling phases;
laboratory pump phases must compensate mixer and cable offsets.
An independently pumped conversion link realizes the Hermitian hopping.
The baseline onsite loss is engineered and can be reprogrammed when
the reservoirs change.

For cold resonant auxiliaries with independent baths and matched
linewidths, assign one auxiliary to each jump vector.
Conversion and pair-creation coupling vectors $u_{\ell,\mu},u_{g,\mu}$ give
$\Gamma_\ell^{\rm prog}=(4/\kappa_b)\sum_\mu
u_{\ell,\mu}u_{\ell,\mu}^\dagger$ and
$\Gamma_g=(4/\kappa_b)\sum_\mu u_{g,\mu}u_{g,\mu}^\dagger$.
Banded Cholesky factors of positive definite range-$r$ matrices have
columns on at most $r+1$ consecutive sites.  Range two therefore needs
auxiliaries coupled to at most three sites and retains the nearest-neighbor
signal drift.  The complete matrix cap includes overlapping channels;
it does not specify pump watts.  All model rates are in units of $J_0$.
Thus $\Omega=0.05$ gives halfwidth $J_0\Omega/(2\pi)=50$ kHz
and full detection bandwidth $100$ kHz.

\begin{table}[t]
\centering
\small
\begin{tabularx}{\linewidth}{@{}>{\raggedright\arraybackslash}p{0.30\linewidth}>{\raggedright\arraybackslash}p{0.27\linewidth}>{\raggedright\arraybackslash}X@{}}
\hline\hline
Quantity & Proposed target & Meaning\\
\hline
Carrier frequencies; temperature & $4$--$8$ GHz; $20$ mK & Design window\\
Rate unit $J_0/(2\pi)$ & $1$ MHz & Converts the model to laboratory units\\
$J_0(t_R,t_L,\gamma)/(2\pi)$ & $(1,0.25,1.2)$ MHz & Internal first-moment drift\\
$J_0(v,c)/(2\pi)$ & $(0.375,0.625)$ MHz & Coherent and baseline bond scales\\
Baseline onsite loss $2\gamma J_0/(2\pi)$ & $2.4$ MHz & Energy decay, not amplitude decay\\
$J_0(\kappa_i,\kappa_o)/(2\pi)$ & $(0.2,0.2)$ MHz & Fixed end ports\\
Band; complete cap $RJ_0/(2\pi)$ & $\pm50$ kHz; $8$ MHz & Same band and both full rate caps\\
Auxiliary energy linewidth $J_0\kappa_b/(2\pi)$ & $80$--$100$ MHz & Matched independent terminations\\
Coupling at each baseline bond end $J_0|g|/(2\pi)$ & $3.536$--$3.953$ MHz & From $c=4|g|^2/\kappa_b$\\
Intrinsic energy loss $J_0\eta/(2\pi)$ & $10$ kHz & Included in complete loss budget\\
\hline\hline
\end{tabularx}
\caption{Proposed microwave design targets.  Dimensionless model rates
are multiplied by $J_0$ and reported after division by $2\pi$.
Optimized jump vectors require different couplings from the baseline;
the representative factors in Sec.~\ref{sec:s_calibrated_resources}
reach $10.361$ MHz.
Frequency placement, unwanted mixing, and pump dynamic range require
an actual circuit design.}
\label{tab:s_experimental_parameters}
\end{table}

\subsection{Finite auxiliaries, intrinsic loss, and the measurement test}
\label{sec:s_experimental_check}

Let $D=\eta I$ be the fixed intrinsic loss matrix.  The complete
loss is $\Gamma_\ell=D+\Gamma_\ell^{\rm prog}$, so implementability
requires $\Gamma_\ell-D\succeq0$ as well as both full rate caps.
For a cold intrinsic bath, the Markov positive branch imposes
$\mathcal N-f\ge y^\dagger D y/(\kappa_i|y_i|^2)$; exact pointwise
saturation is therefore unavailable when $D$ has full rank.

The finite-linewidth response and thermal force noise are given in
Methods, Eqs.~\textup{(36)}
and \textup{(37)}.  Common calibrated drift
and ports, intrinsic loss $D$, and resonant auxiliary linewidth $\kappa_b$
ensure matched responses
even when the reservoir factorizations differ.  The exact two-device
difference is Eq.~\textup{(38)}.
Common intrinsic-loss and thermal offsets cancel from this difference;
finite linewidth still changes the response and frequency weight,
and increasing bath occupation raises each device's absolute noise.
The full signal--auxiliary checks are reported in
Sec.~\ref{sec:s_verification}.

\subsection{Calibrated comparisons and total coupling resources}
\label{sec:s_calibrated_resources}
\begin{figure}[t]
\centering
\includegraphics[width=0.93\textwidth]{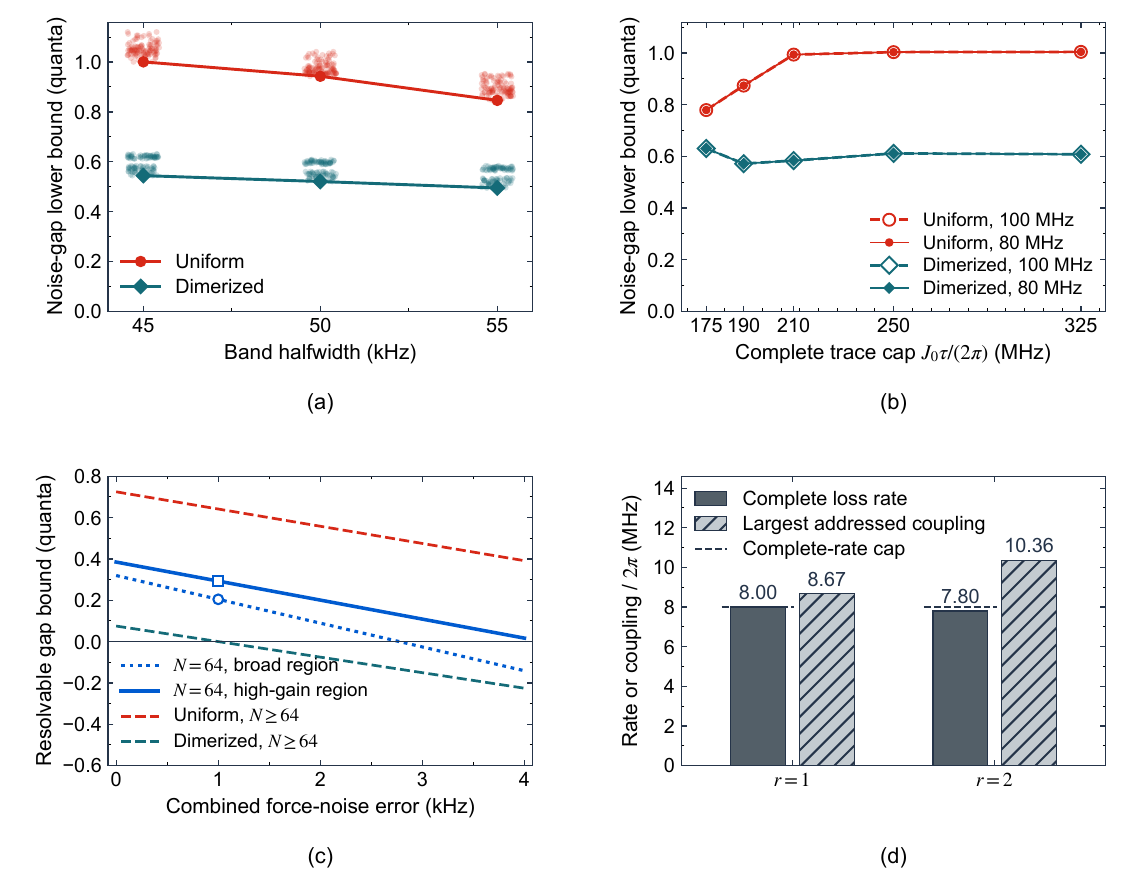}
\caption{Physical comparisons, error tolerances, and resource calibration.
(a) $N=64$ calibrated grid: verified nearest-neighbor lower bound minus
one range-two feasible upper bound, with the reference fixed across
the three halfwidths.  Translucent points show parameter cases; solid
markers show minima, and lines only guide the eye.
(b) Nominal finite-auxiliary devices with $R=8$ and common complete trace
cap $\tau$: all 20 comparisons are shown.  Solid lines with filled markers
denote $J_0\kappa_b/(2\pi)=80$ MHz; dashed lines with open markers, $100$ MHz.
The linewidth curves nearly coincide without any data offset.
Red/circles and teal/diamonds denote uniform and dimerized chains in (a,b).
(c) Resolvable lower bound $\Delta_0-C_Q\delta-2u_{\rm meas}$ versus combined
force-noise error $\delta$, with $u_{\rm meas}=0.05$ per device:
dotted blue, broad continuous uniform $N=64$ region; solid blue, its
high-gain subregion; dashed red/teal, nominal uniform/dimerized Markov
chains for every finite $N\ge64$.  Negative lower bounds do not certify
a resolvable positive gap and do not imply a reversed physical difference.
(d) Representative uniform-chain devices without the trace cap:
solid gray, complete loss rates; hatched gray, largest addressed auxiliary
couplings.  The dashed $8$ MHz cap applies only to complete rates.}
\label{fig:s_robustness}
\end{figure}

The physical optimization enforces $\Gamma_\ell-D\succeq0$,
$\Gamma_g\succeq0$, both complete caps, and the fixed difference $M$.
The calibrated grid below matches the complete spectral caps;
the separate nominal comparisons additionally impose a total trace cap.
For each of the two drifts, multiply all damping, rightward couplings,
and leftward couplings by independently selected factors
$1+\epsilon_\gamma$, $1+\epsilon_R$, $1+\epsilon_L$, with each
$\epsilon\in\{-0.005,0,0.005\}$; this preserves spatial period.
With $\kappa_b\in\{80,100\}$ and $\eta\in\{0,0.02\}$, calibrate 216
devices at $N=64$.  Design one range-two reservoir at $\Omega=0.05$
with cap $7.8$, then hold it fixed at halfwidths $0.045$, $0.05$, and
$0.055$.  The nearest-neighbor benchmark can be optimized separately
at each halfwidth with the larger cap $8$.  At $n_b=0.001$, its verified
lower bound minus the range-two feasible upper bound exceeds $0.84621$
(uniform) and $0.49473$ (dimerized) in all 648 comparisons
[Supplementary Fig.~\ref{fig:s_robustness}(a)].  The largest nearest-neighbor
primal--dual interval is $5.73\times10^{-5}$.
Equation~\textup{(38)} extends the positive signs to
$n_b\in[0,0.02]$.  Parameter and halfwidth coverage here is discrete.

Correcting fixed nominal reference matrices certifies superiority in
only 274 of 648 cases, with smallest lower-bound comparison $-1.74084$.
This failure concerns that reference; the range-two feasible set still
contains the nearest-neighbor set.  A negative comparison lower bound
does not show a negative true optimality gap.  The successful protocol calibrates
before design, then freezes the reservoir across frequency and the
three tested bands.

Methods, Eq.~\textup{(39)}, converts the complete
trace into total squared signal--auxiliary coupling strength,
including the fixed intrinsic-loss offset.

For the finite-auxiliary certificates, it is convenient to use
programmed loss $F_{\rm prog}=\Gamma_\ell-D_{\rm int}$,
with $D_{\rm int}=\eta I$.  The complete positivity and rate conditions are
\begin{equation}
\begin{gathered}
F_{\rm prog},\quad F_{\rm prog}+M+D_{\rm int},\quad
(R-\eta)I-F_{\rm prog},\quad
RI-F_{\rm prog}-M-D_{\rm int}\succeq0,\\
2\operatorname{Tr}F_{\rm prog}+\operatorname{Tr}M+2N\eta\le\tau.
\end{gathered}
\label{eq:s_programmed_loss_constraints}
\end{equation}
Assign positive matrices $A_p,B_p,C_p,D_p$ and trace multiplier
$\xi_p\ge0$ to these constraints.  With the thermal weight
$V=(1+2n_b)\langle|\alpha|^2zz^\dagger\rangle$, the allowed-entry
residual is $E_{\rm prog}=\Pi_r(V-A_p-B_p+C_p+D_p+2\xi_pI)$.
Because $0\preceq F_{\rm prog}\preceq(R-\eta)I$, the residual penalty
is $(R-\eta)\Lambda(E_{\rm prog})$, with $\Lambda$ defined in
main-text Eq.~\textup{(30)}.
The saved code uses the cold weight $Z=\langle|\alpha|^2zz^\dagger\rangle$.
Its programmed-loss objective restores the fixed term
$\operatorname{Tr}[Z(M+D_{\rm int})]$ to obtain the gain-noise bound.
Multiplying this bound and its dual factors by $1+2n_b$ and adding
the common offset $q_{\rm th}$ in Methods gives the reported thermal
noise.  Primal repairs use an anchor with both spectral and trace slack.
This is the same physical optimization as the normalized dual
correction in Methods, not an additional resource assumption.

At common $\kappa_b$, $D$, and $N$, imposing the same total trace cap
therefore imposes the same bound on $\mathcal P_U$.
It does not fix pump watts, which also depend on mixer conversion,
line attenuation, and frequency.  For nominal drifts at $N=64$,
$\eta=0.01$, $n_b=0.001$, and $\Omega=0.05$, add
$\operatorname{Tr}(\Gamma_g+\Gamma_\ell)\le\tau$ to both $R=8$ problems.
The choices $\tau\in\{175,190,210,250,325\}$ and
$\kappa_b\in\{80,100\}$ give 20 comparisons across the two models.
All lower-bound differences exceed $0.57191$
[Supplementary Fig.~\ref{fig:s_robustness}(b)].  For the uniform model at $\tau=175$,
$\kappa_b=80$, nearest-neighbor noise is bounded below by $4.240790$,
while a feasible range-two reservoir is bounded above by $3.461230$;
both use approximately the same complete trace $175$.
Nearest-neighbor primal--dual intervals are at most $4.56\times10^{-7}$.
The range-two intervals can be loose: only feasible upper bounds, not
claims of its exact optimum, enter the separation.

Without the additional trace cap, a representative uniform device at
$\kappa_b=80$, $\eta=0.01$, $\Omega=0.05$, and $n_b=0.001$ has
noise $4.037171$ for $r=1$ and $3.039996$ for $r=2$.
Both implementations use 64 gain and 64 loss auxiliaries; their banded
factors address 254 and 378 signal--auxiliary pairs, respectively.
The largest addressed couplings are $8.666531$ and $10.360989$ MHz,
whereas the maximum complete loss rates are $8.000000$ and $7.800000$
MHz [Supplementary Fig.~\ref{fig:s_robustness}(d)].  A matrix-rate cap does not cap each
individual coupling at the same number.
Their complete traces differ, $213.430888$ versus $321.539906$ MHz;
only the additional trace-constrained comparison above tests matched
total coupling budgets.

\subsection{Continuous working region and measurement tolerances}
\label{sec:s_continuous_device_region}
For the continuous test, take two fixed uniform-chain $N=64$ seeds
designed at $\Omega=0.05$ with cap $7.8$.
Common positive corrections implement $\Delta M$: a diagonal change
$d$ adds on-site gain or loss according to sign; a bond change $d$ uses
gain/loss outer products of weight $|d|/2$ with opposite relative signs.
Their difference gives $d$ without extending the range.
Adding $\eta I$ to both matrices accommodates intrinsic loss, and
coherent conversion is calibrated to the new drift.
At each working point the pumps remain fixed across frequency and bandwidth.
These settings vary with the calibrated parameters; this is not one
uncorrected pump configuration over the whole parameter region.
Only the complete spectral caps are matched here, without a total trace cap.

The continuous box has global, independently varying parameters
\begin{equation}
\begin{gathered}
\gamma\in[1.194,1.206],\quad
t_R\in[0.995,1.005],\quad t_L\in[0.24875,0.25125],\\
\kappa_b\in[80,100],\quad \eta\in[0,0.02],\quad
n_b\in[0,0.02],\quad \Omega\in[0.03,0.07].
\end{gathered}
\label{eq:s_physical_parameter_box}
\end{equation}
The ports remain $\kappa_i=\kappa_o=0.2$ and the complete cap is $8$.
These are uniform-chain parameter errors, not site disorder.
The maximum complete rate is bounded by $7.8445$.
Outward-rounded coefficient intervals and causal Green-function
convolution bound the normalized response on 262,144 parameter cells,
each with 280 frequency intervals of width $0.00025$.
The quadratic-form bound in Eq.~\textup{(40)} is
integrated on each interval; a partial-cell bound covers every halfwidth
in $[0.03,0.07]$.  Thus frequency and bandwidth coverage is continuous.
The resulting device difference exceeds $0.419615$, with whole-band
power gain at least $3.335778$ ($5.23197$ dB).

For the high-gain subregion, restrict each relative drift variation
to $\pm0.00125$, keeping the other intervals unchanged.
Its 4,096 covered cells certify $\Delta_{12}>0.484771$ and
$G(\omega)>16.745077$ ($12.23887$ dB) throughout the band.
Both statements compare the two specified feasible devices, without
asserting continuous optimality relative to every nearest-neighbor
reservoir.

Methods, Eq.~\textup{(41)}, derives the resolvable
gap for matched responses and ports, assuming retained positivity and
complete rate feasibility.  The interval response bounds also give
the required response-energy ceiling $C_Q$.

The broad box has $C_Q<115.258221$, and the high-gain subregion has
$C_Q<92.060749$.  At $\delta=0.001$ and $u_{\rm meas}=0.05$, the respective
residual differences exceed $0.204357$ and $0.292710$ quanta
[Supplementary Fig.~\ref{fig:s_robustness}(c)].  In laboratory units this is a
combined force-noise error of $1$ kHz after division by $2\pi$.
The bound assumes response matching and retained positivity and rate
budgets.  It is not a bound on independent pump-percentage errors.

Temporal stability is checked separately from spatial decay and noise
separation.  Choose a diagonal similarity $S_{\rm sim}$ such that
$A=S_{\rm sim}^{-1}X_\kappa S_{\rm sim}$ is symmetric and negative
definite, and set $x=S_{\rm sim}^{-1}\langle\bm a\rangle$ and
$B=S_{\rm sim}^{-1}(M+D)S_{\rm sim}/2$.
Let $\beta_\ell=(\langle b_{\ell,\mu}\rangle)_\mu$ and
$\beta_g=(\langle b_{g,\mu}^\dagger\rangle)_\mu$ be columns of auxiliary
means.  Their signal feedback is
$u_{\rm fb}=-{\mathrm{i}} S_{\rm sim}^{-1}(U_\ell\beta_\ell+U_g\beta_g)$.
The auxiliary equations give
$\dot u_{\rm fb}=-\kappa_bu_{\rm fb}/2+\kappa_b Bx/2$, and the signal
equation is $\dot x=(A-B)x+u_{\rm fb}$.
Defining the memory correction $\zeta=u_{\rm fb}-Bx$ gives
\begin{equation}
\dot x=Ax+\zeta,\qquad
\dot\zeta=-BAx-(\kappa_b I/2+B)\zeta.
\label{eq:s_memory_dynamics}
\end{equation}
If $-A\succeq\mu I$, $\|A\|\le a$, $\|B\|\le b$, then
$\|x\|^2+\|\zeta\|^2/(ba)$ decreases strictly when
$\mu(\kappa_b/2-b)>ba$.
For completeness, differentiating
$\mathcal E=\|x\|^2+\|\zeta\|^2/(ba)$ along these memory equations gives
\begin{equation}
\dot{\mathcal E}\le-2\mu\|x\|^2+4\|x\|\|\zeta\|
-\frac{2(\kappa_b/2-b)}{ba}\|\zeta\|^2.
\label{eq:s_memory_energy_detail}
\end{equation}
The associated two-variable quadratic form is strictly negative
when $\mu(\kappa_b/2-b)>ba$.  This supplies temporal stability
independently of the spatial-root bounds used for noise tails.

For the broad uniform box, $\mu=0.189$, $a=2.311$, and
$b=2.794184$ give margin above $0.5745$ at $\kappa_b=80$.
This is a stability condition for every finite $N\ge64$;
auxiliary combinations absent from $u_{\rm fb}$ decay at $\kappa_b/2$.
The finite-auxiliary noise-region certificate itself remains at $N=64$.

For measurement, keep one pump configuration over the band.  A vector
network analyzer measures complex transmission and reflection; local
weak probes calibrate coupling vectors, complete rates, and intrinsic
loss.  With the coherent tone off, retain the ports and terminate the
input in a calibrated cold load.  Output isolators, a
high-electron-mobility transistor (HEMT) amplifier, and an
in-phase/quadrature (I/Q) receiver measure the noise.
A reference-load temperature sweep calibrates receiver gain, loss,
added noise, and the heterodyne vacuum offset \cite{LecocqConversion2016}.
Subtract the vacuum floor, divide by measured power gain, and average
over the same band for both ranges; Eq.~\textup{(41)}
sets the required uncertainty.  Pump noise, finite-temperature ports,
unequal auxiliary linewidths or detunings, and compression need separate
characterization.  A shorter demonstrator also needs its own high-gain
parameter choice and reservoir optimization.

\FloatBarrier

\section{Computational verification and data availability}
\label{sec:s_verification}

\subsection{Finite-matrix and all-length checks}
CVXPY \cite{DiamondBoyd2016} and Clarabel \cite{GoulartChen2024}
supply candidate primal matrices and dual factors.  Exact rational
matrix products and polynomial integration then enclose the optima
using feasible primal objectives and residual-corrected dual bounds,
independently of solver status.  For the all-length results, these
checks also verify the reserved bridge budget, Schur ellipse, geometric
tail, and continuous interval coverage in Sec.~\ref{sec:s_size_uniform}.
Finite-chain resolvents, spectra, and independent noise integrals provide
floating-point cross-checks.  The all-length claims follow from the
compatible boundary response and controlled extension.
The reported gaps and targets hold for the two nominal chains in
Eq.~\eqref{eq:s_two_jacobi_models} at $N\ge64$; general disorder,
interactions, multiorbital mixing, and sharp universal range or
bandwidth scaling remain open.  Solver settings, entry points, and
source and matrix hashes are recorded with the data.

The unrestricted pointwise floor and the qualitative limitation of one
bath optimized at multiple detunings \cite{LauClerk2018}, amplifier noise
limits \cite{Caves1982,Clerk2010}, reservoir-engineered quantum-limited
amplification and directionality
\cite{MetelmannClerk2014,MetelmannClerk2015,RamosDirectional2021}, and
topological parametric amplification \cite{PeanoEtAl2016,GomezLeonEtAl2023}
are established results.  Long-range dissipative couplings also produce
multimode topological amplification \cite{VegaEtAl2025}.
The present additions concern conditional adjacent-cell attainment,
shared-band costs at matched drift, ports, and complete rate caps, and the calibrated-device
and noise-error bounds.

\subsection{Checks for calibrated devices}
The physical comparisons retain four PSD dual factors and a trace
multiplier when required; an interior primal repair and dual residual
correction give the bounds.  The 193-point Gauss--Legendre averages
agree with 97-point checks within $1.14\times10^{-13}$.
Independent tests cover 192 exact dyadic PSD constraints, eight full
signal--auxiliary comparisons, and 48 response samples at 70-digit
precision.  Full and reduced responses and thermal noises agree within
relative error $1.15\times10^{-14}$; spectra use diagonal balancing.
Continuous-region checks follow Sec.~\ref{sec:s_continuous_device_region},
and nominal all-length noise errors use rational sums and tail bounds.

The extended 58,320-case scan at $N=64,96,128$ gives positive feasible
device differences.  All samples enter the noise-difference and sign
checks, including 864 with gain below unity; those 864 are not
amplifier operating points.  These scans do not establish continuous
optimality.  Failed frozen-reference tests are retained.  Unequal
auxiliary linewidths or detunings, independent pump errors, disorder,
thermal ports, compression, and actual pump watts remain unverified.

\FloatBarrier

\subsection*{Data and code availability}
Data and calculation code for all results are public in repository
version 1.3.1 \cite{LiReservoirCode2026}.  Supplementary Data contains
the core code, inputs, configuration and result data; the full
calibrated-reference and matrix-witness banks are supplied in documented
public archives.  The English guide at
\path{calculations/robustness_20261007/README.md} maps claims to saved
artifacts by model, size, budget and calibration scope.  It gives
archive checksums, extraction instructions, commands and success
markers, and distinguishes saved-result checks from optimizer reruns.
Frozen-reference and below-unity-gain controls are retained.
No experimental data were collected.